\documentclass[12pt]{article}

\usepackage{float}
\usepackage[makeroom]{cancel}
\usepackage{color}
\usepackage{graphicx}
\usepackage{amsmath}
\usepackage{amssymb}
\usepackage{xspace}
\usepackage[small]{subfigure}
\usepackage[numbers,compress]{natbib}
\usepackage[hyperfootnotes=false]{hyperref}
\usepackage{tcolorbox}

\usepackage{framed}
\usepackage{physics}
\usepackage{tensor}

\newlength{\fighskip} \fighskip=2pt
\newlength{\figvskip} \figvskip=3pt

\newcommand*{\figbox}[2]{{
  \def\figscale{#1}
  \def\arraystretch{0.8}
  \arraycolsep=0pt
  \begin{array}{c}
    \vbox{\vskip\figscale\figvskip
      \hbox{\hskip\figscale\fighskip
        \includegraphics[scale=\figscale]{#2}}}
  \end{array}}}

\usepackage{mciteplus} 
\usepackage{dcolumn}
\usepackage{bm}
\usepackage{verbatim}
\usepackage{amscd}
\usepackage{amsfonts}
\usepackage{setspace}
\usepackage{amsthm}
\usepackage{enumerate}
\usepackage{mathtools}

\theoremstyle{plain}
\newtheorem{claim}{Claim}

\theoremstyle{plain}

\theoremstyle{plain}

\theoremstyle{remark}

\theoremstyle{observation}

\theoremstyle{definition}

\theoremstyle{corollary}

\theoremstyle{definition}

\theoremstyle{definition}

\theoremstyle{result}
\newtheorem{definition}{Definition}
\theoremstyle{assumption}

\theoremstyle{definition}

\theoremstyle{problem}

\theoremstyle{fact}

\newcommand{\BY}[1]{{\textcolor{purple}{#1}}}
\newcommand{\DL}[1]{{\textcolor{orange}{#1}}}

\usepackage{authblk}
\title{
\bf %\boldmath 
Chiral Color Code
: Single-shot error correction for exotic topological order
}

\author[1,2]{Dongjin Lee
}
\author[1]{Beni Yoshida
}
\affil[1]{\em \small Perimeter Institute for Theoretical Physics, Waterloo, Ontario N2L 2Y5, Canada}
\affil[2]{\em \small Department of Physics and Astronomy, University of Waterloo, Waterloo, Ontario N2L 3G1, Canada}
\affil[ ]{\textit {\href{mailto:dlee@perimeterinstitute.ca}{dlee@perimeterinstitute.ca}, \hspace{3pt} \href{mailto:byoshida@perimeterinstitute.ca}{byoshida@perimeterinstitute.ca}}}

\date{}

\begin{document}
\maketitle

\begin{abstract}
We present a family of simple three-dimensional stabilizer codes, called the chiral color codes, that realize fermionic and chiral topological orders. In the qubit case, the code realizes the topological phase of a single copy of the fermionic toric code. For qudit systems with local dimension $d$, the model features a chiral parameter $\alpha$ and realizes 3D topological phases characterized by $\mathbb{Z}_d^{(\alpha)}$ anyon theories with anomalous chiral surface topological order. 
On closed manifolds, the code has a unique ground state after removing bulk transparent fermions or bosons. 
Furthermore, we prove that the bulk is short-range entangled (for odd $d$, coprime $\alpha$) by constructing an explicit local quantum channel that prepares the ground state. 
The chiral color codes are constructed within the gauge color code, and hence inherit its fault-tolerant features: they admit single-shot error correction and allow code switching to other stabilizer color codes. 
These properties position the chiral color codes as particularly useful platforms for realizing and manipulating fermions and chiral anyons. 

\end{abstract}

\tableofcontents

\newpage

\section{Introduction}

Topological quantum error-correcting codes provide a natural platform for fault-tolerant quantum computing by protecting quantum information against local physical noises. 
Prominent examples include the toric code~\cite{Kitaev97} and the color code~\cite{Bombin_2006}, which realize bosonic $\mathbb{Z}_2$ topological orders with abelian anyonic excitations.
In conventional settings, physical errors are corrected by projectively measuring local syndrome operators that detect anyon locations, followed by annihilating these anyons in a manner that avoids introducing logical errors~\cite{Dennis_2002}. 
However, if the syndrome measurements themselves are faulty, they may induce effective non-local errors which potentially lead to logical failures. 
In fact, all known 3D stabilizer codes are susceptible to imperfect syndrome measurements and thus require repeated rounds of measurements in every step of error correction, leading to significant overhead.

A key breakthrough in this direction is the discovery of \emph{single-shot error correction}, which enables error correction using only a single round of noisy syndrome measurements~\cite{Bombin:2014ksv}. 
Namely, it has been demonstrated that a certain subsystem code, the 3D gauge color code~\cite{Bombin:2015tpp}, overcome this limitation due to faulty syndrome measurements as the code possesses local redundancy relations, often called meta-checks, among its gauge operators.
This allows for error correction of noisy syndrome values in a single step and keeps the effective physical error rates suppressed even under imperfect syndrome measurements.  
Furthermore, the single-shot error-correction capability in the 3D gauge subsystem color code enables fault-tolerant code switching between the 3D stabilizer color code and the 2D stabilizer color code, enabling fault-tolerant implementation of a universal gate set~\cite{Bombin:2016aoq}. 
The single-shot error-correction capability may significantly reduce the overhead and provide an appealing route toward scalable quantum computation.

Despite these major advances in quantum error correction, the known landscape of topological codes capable of single-shot error correction has remained largely limited to bosonic and non-chiral phases, in practice, to variants of the $\mathbb{Z}_2$ toric code and its direct generalizations~\cite{Kubica:2021gtj, Brown_2016, Watson2015, Bridgeman2025, Stahl_2024}.
By contrast, many physically realized topological orders, such as those in fractional quantum Hall systems, are chiral, and where emergent fermionic excitations can arise even from purely bosonic microscopic degrees of freedom. 
Also, the underlying physical mechanism behind single-shot quantum error correction has remained unclear. 
A central open question, from the perspective of both condensed matter physics and quantum information science, has been to construct simple, exactly solvable stabilizer models that realize fermionic or chiral topological orders, while retaining the fault-tolerance properties such as single-shot error-correction capability.

\subsubsection*{Main result}

In this paper, we present a family of simple 3D stabilizer codes, called the \emph{chiral color codes}, that realize fermionic and chiral topological orders. 
We first introduce the XYZ color code, a qubit version of the chiral color code. 
This stabilizer code is defined on a four-colorable lattice, and can be viewed as a deformation of the color code where Pauli operators $X$, $Y$, and $Z$ in face stabilizers depend on face color labels.
\footnote{
A version of the XYZ color code has been previous considered by Kim for a particular lattice in~\cite{Kim2010}. 
However, its physical properties remained largely unexplored, including the presence of fermionic excitations. 
That said, Kim already argued that his model has finite temperature topological order while it serves only as a classical memory, which is consistent with the recent finding on the fermionic toric code~\cite{Zhou:2025bal}.
}
We show that the XYZ color code realizes the topological phase of a single copy of the fermionic toric code~\cite{Levin_2003, Chen:2018nog} while supporting fermionic anomalous surface topological order.
We then generalize the XYZ color code to qudits and construct the chiral color code. 
For qudit systems with local dimension $d$, the model features a chiral parameter $\alpha$ where Pauli operators $X$, $Z$, and $XZ^{\alpha}$ in face stabilizers depend on face color labels. 
The model realizes 3D topological phases characterized by $\mathbb{Z}_d^{(\alpha)}$ anyon theories with anomalous chiral surface topological order. 

As the name suggests, the chiral color codes are constructed within the gauge color code, and hence inherit its fault-tolerant features, namely they admit single-shot error correction and allow code switching to other stabilizer color codes where non-Clifford gates can be transversally implemented. 
This capability also enables single-shot preparation of a 2D chiral $\mathbb{Z}_d^{(\alpha)}$ mixed state $\rho$ that emerges on the boundary of the chiral color code.
To our knowledge, this is the first construction that realizes fermionic and chiral topological orders while admitting single-shot error correction.

\subsubsection*{Side results}

A hallmark property of a 3D $\mathbb{Z}_d^{(\alpha)}$ topological phase is that the bulk is short-range entangled after removing transparent bulk boson or fermion. 
This fact has been confirmed in a few specific instances~\cite{Haah_2022, Shirley:2022lhu, Haah:2019fqd, Bauer_2023} of the Walker-Wang type models~\cite{Walker:2012mcd}, but unified proofs have been missing.
%\DL{(I became aware of this paper \cite{Bauer_2023} (I think Tyler mentioned to us before), which disentangles WW model with modular input anyon being Drinfield center, meaning that it has a gappable boson (this includes $\mathbb{Z}_9^1$ case). It also presented disentangling unitaries with fermionic ancillas when the input anyon theory is in the Witt class of Ising UMTC, but I am not sure how much this case overlaps with our construction.)}
Here, we prove that the bulk of the chiral color code is indeed short-range entangled for odd $d$ and coprime $\alpha$, with a unique ground state on closed manifolds.
(For even $d$ or non-coprime $\alpha$, the model contains a bulk fermion or boson, and thus is long-range entangled.)
Namely, we explicitly construct a local quantum channel 
%which can be regarded as quantum cellular automaton, % 
that prepares the bulk ground state.
Hence, the long-standing belief, concerning bulk short-range entanglement in the Walker-Wang type models, can be rigorously confirmed for a wide variety of chiral topological phases in a unified footing within the framework of the chiral color code. 

Given the complexity of the Walker-Wang type constructions, the remarkable simplicity of our construction, based on simple stabilizer constructions, will be of great value for both conceptual understanding of corresponding topological phases as well as their physical realizations. 
Indeed, the simplicity of the color code like models often manifest hidden symmetry properties, such as emergence of SPT-like defects~\cite{Yoshida:2015boa, Yoshida:2015cia}. 
We thus expect that our construction will find applications far beyond the original scope of single-shot error correction. 
In fact, our construction already suggests simple realization of chiral mixed states with $\mathbb{Z}_d^{(\alpha)}$ anyons.
Also, since experimental realizations of fermions and chiral anyons may be of interest in near-term device applications, we present small-size examples ($8$ and $15$ qudits) in section~\ref{sec:XYZ_surface}. 

Another perspective on our construction comes from the viewpoint of SPT phases with higher-form symmetries.
It is already known that the gauge color code possesses emergent $1$-form symmetries~\cite{Kubica:2018lhn}, and that they play a central role in enabling single-shot error correction~\cite{Roberts:2016lvf}.
Our models provide yet another concrete realization of this phenomenon: The XYZ and chiral color codes can be viewed as SPT phases protected by $1$-form symmetries, where the anomaly of the surface chiral order is canceled by the short-range entangled bulk.
The single-shot error-correction capability is closely tied with the fault-tolerance of measurement-based quantum computation (MBQC) resource states~\cite{RBH, Roberts:2016lvf, Yoshida:2015cia, Kubica:2018lhn}.
Our result suggests that the chiral color code serves as a MBQC resource state where quantum computations are performed with chiral anyons. 
This perspective is briefly discussed in section~\ref{sec:chiral}. 
A relevant development can be also found in a previous work for abelian non-chiral anyon theories~\cite{Bauer2025}.

While conventional quantum computation is formulated within bosonic qubit systems, fermionic quantum computation has particular advantage over the bosonic counterpart~\cite{Bravyi_2002}. 
In exploring a hybrid quantum computing architecture utilizing both advantages, the XYZ color code or its variants may serve as a useful conversion platform between bosonic and fermionic systems.
\footnote{For instance, one may fermionize a part of the XYZ color code to obtain a hybrid quantum error-correcting code.}
Furthermore, the fault-tolerant code switching capability enables us to implement logical $T$-gates on fermionic degrees of freedom via the XYZ color code.
\footnote{In principle, a logical $T$ gate in the fermionic toric code can be also implemented by sweeping fermionic SPT phase, i.e. $p + ip$ superconductor~\cite{Barkeshli:2023bta}. The XYZ color code suggests an alternative route for implementing non-Clifford gates in the fermionic toric code.}
Similarly, the chiral color code may serve as a useful platform to convert chiral anyons into bosonic degrees of freedom and vice versa. 
The capability of implementing non-Clifford gates on chiral anyons via code switching will be useful in such setups. 
Ultimately, such a fault-tolerant protocol in the chiral color code may suggest a protocol to implement non-Clifford gates in naturally arising topological phases, such as in fractional quantum Hall systems.

In principle, arbitrary abelian anyon theories can be constructed from multiple copies of $\mathbb{Z}_d^{(\alpha)}$ anyon theories after appropriate boson condensations, consistent with the Witt class classification~\cite{Haah:2019fqd}.
As examples, in appendix~\ref{appendix:condensation}, we demonstrate how to obtain the chiral semion and the three fermion theory from $\mathbb{Z}_4^{(1)}$ anyon theories via boson condensation.
A significant open question concerns non-abelian generalizations, which is left for future work.

\subsubsection*{Plan of the paper}

This paper is organized as follows.
In section~\ref{sec:single-shot}, we review the notion of single-shot error-correction. 
In section~\ref{sec:XYZ}, we introduce the XYZ color code and discuss its fermionic excitations. 
In section~\ref{sec:XYZ_surface}, we study the surface fermionic topological order. 
In section~\ref{sec:chiral}, we introduce the chiral color code and show that it realizes the $\mathbb{Z}_d^{(\alpha)}$ anyon theory. 
In section~\ref{sec:chiral_SRE}, we prove that the chiral color code is short-range entangled when defined on a closed manifold. 
In section~\ref{sec:chiral_surface}, we study the anomalous surface chiral topological order in the chiral color code. 

%{sec:single-shot}
%{sec:XYZ}
%{sec:XYZ_surface}
%{sec:chiral}
%{sec:chiral_SRE}
%{sec:chiral_surface}
%{sec:outlook}

\section{Single-shot error correction in gauge color code}\label{sec:single-shot}

We provide a brief review of single-shot error correction and its implementation in the 3D gauge color code. 

\subsection{Single-shot error correction}

\subsubsection*{- Physical errors}

\begin{figure}
\centering
\raisebox{\height}{a)\hspace{10pt}}\raisebox{-0.85\height}{\includegraphics[width=0.2\textwidth]{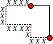}}
\hspace{10pt}
\raisebox{\height}{b)\hspace{10pt}}\raisebox{-0.85\height}{\includegraphics[width=0.23\textwidth]{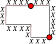}}
\hspace{10pt}
\caption{2D toric code with physical errors. 
Red dots represent $m$ anyons. 
a) Dotted lines represent the actual $X$-errors while solid lines represent the minimal weight error-correction path. 
b) The total Pauli-$X$ action after the error-correction. 
They are likely to form contractible closed loops of Pauli-$X$ operators.
}
\label{fig_2DTC}
\end{figure}

Let us begin by recalling how \emph{physical errors} are corrected in the 2D toric code~\cite{Dennis_2002}:
\begin{align}
\mathcal{S}_{\text{2DTC}} = \langle A_v, B_p \rangle  
\qquad  A_v = \figbox{2.5}{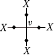}, \qquad B_p = \figbox{2.5}{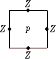}\ .
\end{align}
Assume that the system is subject to dephasing noises by local Pauli $X$ operators, each occurring with a small probability $p$. 
Such errors create pairs of $m$ anyons by violating $Z$-type plaquette stabilizers. 
A natural error-correction strategy is to pair up the $m$ anyons along paths of minimal total weights and annihilate them (Fig.~\ref{fig_2DTC}(a)).  
This strategy is effective because stochastic $X$ errors are unlikely to generate widely separated anyon pairs.
Namely, the probability of forming a string of Pauli $X$ operators of length $\ell$ scales as $\sim p^{\ell}$. 
While the error-correction path may differ from the actual $X$-error path, the total $X$-action on the system forms closed loops (Fig.~\ref{fig_2DTC}(b)).
It is very unlikely to find non-contractible loops of $X$ operators as they require $O(L)$-weight $X$-errors that are exponentially suppressed. 
As such, the residual $X$-action after the error correction is trivial on the codeword subspace. 

\subsubsection*{- Faulty syndrome measurements}

\begin{figure}
\centering
\raisebox{\height}{a)\hspace{10pt}}\raisebox{-0.85\height}{\includegraphics[width=0.24\textwidth]{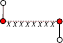}}
\hspace{10pt}
\raisebox{\height}{b)\hspace{10pt}}\raisebox{-0.85\height}{\includegraphics[width=0.27\textwidth]{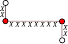}}
\hspace{10pt}
\caption{2D toric code with physical errors and faulty syndromes (shown by white dots). 
a) The error-correction operation (shown in a solid line) that may mix up faulty syndrome and $m$ anyon. 
b) The total $X$ action after the error-correction. An open $X$-string connecting faulty syndrome locations is introduced as a residual error.  
}
\label{fig_2DTC_faulty}
\end{figure}

Next, let us look at the effect of \emph{faulty syndrome measurements} in the 2D toric code where the outcome of measuring each $B_{p}$ may be flipped with some small probability $q$~\cite{Dennis_2002}.  
Such errors effectively create or annihilate anyonic excitations at each plaquette with probability $q$, making it impossible to reliably determine whether a violation of $B_{p}$ is due to a physical $X$ error or a faulty measurement. 
To gain some intuition, consider a situation in which both physical errors and measurement errors are present, as in Fig.~\ref{fig_2DTC_faulty}(a).  
The decoder may mistakenly pair up a physical anyon with a fictitious anyon created by a faulty measurement, or may pair up fictitious anyons. 
This results in an incorrect correction path, leading to a string of Pauli $X$ operators connecting two faulty syndrome locations as in Fig.~\ref{fig_2DTC_faulty}(b). 
Such residual errors can be non-local as the probability of generating a pair of fictitious $B_p$ violations separated by a distance $\ell$ scales as $\sim q^2$, independent of $\ell$.

Thus, even rare faulty syndrome measurements can effectively introduce high-weight physical errors, unless syndromes are measured multiple times to accurately infer the true error configuration~\cite{Dennis_2002}.
The key question is whether reliable error correction is possible from a single round of noisy syndrome measurements i.e. whether the code admits \emph{single-shot error correction}. 

\subsubsection*{- Meta-check conditions}

This issue can be partially mitigated in the 3D toric code for $X$ errors (though not for $Z$ errors), due to redundancy among the plaquette stabilizers $B_p$:
\begin{align}
\mathcal{S}_{\text{3DTC}} = \langle A_v, B_p \rangle,  
\qquad   A_v = \figbox{3.5}{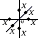},\qquad B_p = \figbox{2.5}{fig_2DTC_P}. 
\end{align}
Observe that the product of the six plaquette stabilizers living on the faces of a single cubic cell yields the identity operator:
\begin{align}
\prod_{p \in \text{cube}} B_{P} = \figbox{3}{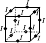}.
\end{align}
As a result, in the absence of faulty measurements, the measured syndrome values $b_p^{(Z)} = \pm 1$ must satisfy a consistency condition, which is commonly called a \emph{meta-check}: 
\begin{align} 
\prod_{p \in \text{cube}} b_p^{(Z)} = 1. 
\end{align} 

\begin{figure}
\centering
\raisebox{\height}{a)\hspace{10pt}}\raisebox{-0.85\height}{\includegraphics[width=0.22\textwidth]{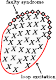}}
\hspace{10pt}
\raisebox{\height}{b)\hspace{10pt}}\raisebox{-0.85\height}{\includegraphics[width=0.25\textwidth]{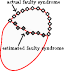}}
\hspace{10pt}
\raisebox{\height}{c)\hspace{10pt}}\raisebox{-0.85\height}{\includegraphics[width=0.21\textwidth]{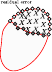}}
\hspace{10pt}
\caption{3D toric with physical errors and faulty syndromes.
a) Closed loops of $B_{p}$ excitations that are partly erased and become open strings due to faulty syndromes (shown in white dots).  
b) Estimations of faulty syndrome locations (shown in filled dots). 
c) Residual errors after the error-correction. 
}
\label{fig_3DTC_faulty}
\end{figure}

To better characterize the error configuration, it is useful to move to the dual lattice $\widetilde{\mathcal{L}}$, where vertices correspond to cubes of the original lattice $\mathcal{L}$. 
Measured syndrome values $b_p^{(Z)} = \pm 1$, associated with plaquettes $p$ in the original lattice $\mathcal{L}$, are then assigned to edges $\tilde{e}$ of the dual lattice $\widetilde{\mathcal{L}}$.
Namely, the meta-check constraints take the form of $\mathbb{Z}_2$ lattice gauge theory conditions 
\begin{align}
\prod_{\tilde{e} \in \text{star}(\tilde{v})} b_{\tilde{e}}^{(Z)} = 1 \qquad (\text{dual lattice } \widetilde{\mathcal{L}}).
\end{align}
This suggests that configurations of valid syndrome values form closed loops in the dual lattice. 
%To better characterize the error configuration, it is useful to move to the dual lattice, where vertices correspond to cubes of the original lattice. 
%The plaquette stabilizer $B_p$ in a dual lattice can be expressed as
%\begin{align}
%B_{p} =  \red{add equation}.
%\end{align}
%Measured syndrome values $s_p^{(Z)} = \pm 1$, associated with plaquettes in the original lattice $\mathcal{L}$, are then assigned to edges of the dual lattice $\widetilde{\mathcal{L}}$.
%Namely, the meta-check constraints take the form of $\mathbb{Z}_2$ lattice gauge theory constraints: 
%\begin{align} 
%\prod_{\tilde{e} \in \text{star}(v)} s_{\tilde{e}}^{(Z)} = \red{add equation} =  1, \qquad (\text{dual lattice}).
%\end{align} 
%This suggests that configurations of valid syndrome values form closed loops (in the absence of faulty syndromes).
Closed loops of $B_{p}$ excitations can be created by Pauli $X$ operators acting on an open membrane as shown in Fig.~\ref{fig_3DTC_faulty}(a). 
For sufficiently small error rate $p$, stochastic $X$ errors tend to generate small-size loops. 
Hence a natural error-correction strategy is to shrink closed loops and annihilate them by applying Pauli $X$'s in their interior.  \\

\subsubsection*{- Correcting faulty syndromes}

Now we discuss how to circumvent the effect of faulty syndromes~\cite{Bombin:2014ksv}. 
Since faulty measurements may violate the meta-check conditions, configurations of measured syndrome values may form open strings as shown in Fig.~\ref{fig_3DTC_faulty}(a). 
The central idea is to correct syndrome values by using the meta-checks. 
While it is not possible to fully determine the locations of faulty syndromes, one can identify the endpoints of open strings formed by faulty sundromes.
When the faulty measurement rate $q$ is sufficiently small, faulty syndromes are unlikely to generate open strings with widely separated endpoints. 
The correction strategy for faulty syndromes then is to connect the endpoints of open strings via minimal paths, interpreting these paths as possible locations of faulty measurements as shown in Fig.\ref{fig_3DTC_faulty}(b).
One then attempts to eliminate closed loops of $B_{p}$ by using the corrected syndrome values. 

Since the corrected syndrome values and the real syndrome values may differ, there can be residual errors as shown in Fig.~\ref{fig_3DTC_faulty}(c). 
The crucial point is that the \emph{residual errors remain low-weight} as the endpoints of open strings are unlikely to be widely separated.
Consequently, the system stays close to the original state up to low-weight errors.

\subsubsection*{- Single-shot error correction}

As seen above, single-shot error correction refers to the ability to perform reliable error correction from a single round of noisy syndrome measurements.
The key caveat is that one cannot fully eliminate all measurement errors. Instead, the goal is to ensure that the resulting residual effective errors remain local (i.e., low weight), so that subsequent quantum operations can still be performed fault-tolerantly.

Formally, single-shot error correction can be characterized as follows~\cite{Bombin:2014ksv, Campbell_2019, Kubica:2021gtj}:

\begin{definition}
\emph{[\textbf{Single-shot error correction}]} A quantum code is said to admit single-shot error correction if, for sufficiently small measurement error rate $q < q_*$, the residual error is $\tau$-bounded.
Here, $\tau$-bounded means that an error of weight $w$ occurs with probability at most $\text{Prob} \leq O(\tau^w)$. 
\end{definition}

Conventionally, we require $\tau(q)$ to be sufficiently small and satisfy $\tau(q) \rightarrow 0$ as $q \rightarrow 0$, see~\cite{ Kubica:2021gtj, Bombin:2014ksv} for details.

Unfortunately, the 3D toric code does not admit single-shot error correction for $Z$ errors as vertex stabilizers $A_v$ are not equipped with meta-check conditions (except a trivial relation of $\prod_v A_v = I$).
It is well known that the 4D toric code (loop only version) admits single-shot error-correction.
To the best of our knowledge, there is no known example of 3D stabilizer codes with single-shot error-correction capabilities. 
This (conjectured) impossibility of single-shot error correction for 3D stabilizer codes is conventionally tied with the (conjectured) impossibility of 3D finite temperature quantum memory~\cite{Bombin:2014ksv, Bravyi:2011zvu, Siva:2016fzs, Brown:2014idi}, although no formal proof exists. 
In the next two subsections, we recall how to circumvent this challenge by using 3D subsystem codes. 

\subsection{Topological color codes}

Next, we recall the construction of the 2D and 3D (stabilizer) color code.

\subsubsection*{- Colorable lattice}

\begin{figure}
\centering
\raisebox{\height}{a)\hspace{10pt}}\raisebox{-0.85\height}{\includegraphics[width=0.4\textwidth]{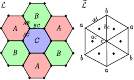}}
\hspace{10pt}
\raisebox{\height}{b)\hspace{10pt}}\raisebox{-0.85\height}{\includegraphics[width=0.45\textwidth]{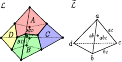}}
\hspace{10pt}
\caption{Colorable lattice $\mathcal{L}$ and dual lattice $\widetilde{\mathcal{L}}$. Black dots represent qubits.  
a) A $2$-colorable lattice. $2$-cells (plaquettes) have colors ${A, B, C }$ and $1$-cells (edges) have colors ${AB, BC, AC }$.
b) A $3$-colorable lattice. $3$-cells (volumes) have colors ${A, B, C, D }$, $2$-cells (plaquettes) have colors ${AB, AC, AD, BC, BD, CD }$, and $1$-cells (edges) have colors ${ABC, ABD, ACD, BCD }$.
}
\label{fig_lattice}
\end{figure}

Let us begin by recalling some basic notations to define (stabilizer) color codes. 
Given a $d$-dimensional lattice ${\cal L}$, we denote the set of its $i$-dimensional cells by $\Delta_i $. 
We will focus on a particular family of $D$-dimensional lattices known as \emph{$D$-colorable lattices}, which satisfy two key conditions: \emph{$(D+1)$-valency} and \emph{$(D+1)$-colorability}~\cite{Bombin:2006rk, kubica2018abcs}. 
\begin{enumerate}
\item A lattice is \emph{$(D+1)$-valent} if each of its vertices is incident to exactly $D+1$ edges. 
\item A lattice is \emph{$(D+1)$-colorable} if each $D$-cell can be assigned one of $(D+1)$ distinct colors such that no two adjacent $D$-cells share the same color.
\end{enumerate}
Illustrative examples for $D=2,3$ are depicted in Fig.~\ref{fig_lattice}.

We will use uppercase letters ${A, B, C, \dotsb}$ to label the colors assigned to $D$-cells.
A color can also be associated with a $d$-dimensional cell (for $0 \leq d \leq D$) based on the set of $(D+1-d)$ adjacent $D$-cells that contain it. 
For $D=3$, possible color labels are:
\begin{align*}
&\text{$3$-cells (volumes): ${A, B, C, D }$} \\
&\text{$2$-cells (faces): ${AB, AC, AD, BC, BD, CD }$}\\ 
&\text{$1$-cells (edges) : ${ABC, ABD, ACD, BCD }$ }
\end{align*}
We also consider the dual lattice $\widetilde{\mathcal{L}}$, where lowercase letters ${a, b, c, \dotsb}$ are used to label colors, to distinguish them from the primal lattice. 
Namely, vertices ($0$-cells) in $\widetilde{\mathcal{L}}$ carry color labels ${a, b, c, \dotsb}$ with neighboring vertices having different colors (Fig.~\ref{fig_lattice}).

Given a $D+1$-colorable lattice $\mathcal{L}$, a single qubit is placed at each vertex of the lattice, i.e., each $0$-cell in $\Delta_0$ (see Fig.~\ref{fig_lattice}). 
For a given $i$-cell $\delta \in \Delta_i$, it will be convenient to define the action of a Pauli operator $O \in \{X,Y,Z\}$ on all qubits on vertices of $\delta$ by
\begin{align}
X(\delta) \equiv \prod_{v \in \delta} X_v, \qquad Y(\delta) \equiv \prod_{v \in \delta } Y_v, \qquad Z(\delta) \equiv \prod_{v \in \delta } Z_v \ .
\end{align}

\subsubsection*{- 2D and 3D color code}

We are now ready to define the stabilizer color code. 

\begin{definition}\emph{[\textbf{2D color code}]}
The stabilizer group of a 2D color code is given by plaquette operators with Pauli $X$ and Pauli $Z$: 
\begin{eqnarray*}
\mathcal{S}_{\text{2DCC}} &=& \langle X(f), Z(f) |f \in \Delta_2\rangle \  \\
&=&  \left\langle \figbox{0.36}{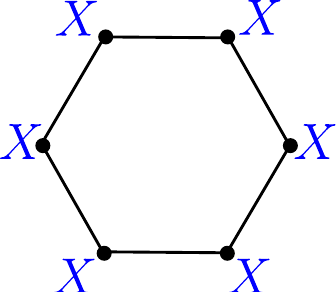}\ , \  \figbox{0.36}{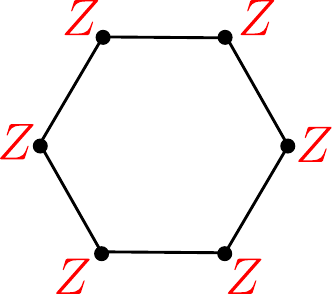}  \right\rangle
\end{eqnarray*}
where we used hexagons to schematically represent plaquettes.
\end{definition}

Stabilizers $X(f)$ and $Z(f')$ commute with each other since neighboring $f$ and $f'$ always share two qubits. 
A 2D topological color code can be mapped to two copies of 2D toric codes using a finite-depth circuit~\cite{Kubica:2015mta}. 
Logical Pauli operators are given by non-contractible $X$ and $Z$ string operators, which connect faces with the same color. 

\begin{definition}\emph{[\textbf{3D color code}]}
The stabilizer group of a 3D color code is given by plaquette and volume operators: 
\begin{eqnarray*}
\mathcal{S}_{\text{3DCC}} &=& \langle X(c), Z(f) | f \in \Delta_2, c \in \Delta_3 \rangle \  \\
&=&  \left\langle \figbox{2.5}{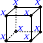}\ , \ \ \figbox{0.36}{fig_2D_Z.pdf}  \right\rangle 
\end{eqnarray*}
where we used cubes to schematically represent volumes. 
\end{definition}

Stabilizers $X(c)$ and $Z(f)$ commute with each other since neighboring $c$ and $f$ always share two qubits. 
A 3D color code can be mapped to three copies of 3D toric code using a finite-depth circuit~\cite{Kubica:2015mta}. 
The code has both string-like and membrane-like logical operators. 
String-like logical operators are given by non-contractible Pauli $Z$ strings connecting Pauli $X$ volume stabilizers with the same color. 
Membrane-like operators are given by non-contractible Pauli $X$ membranes consisting of plaquettes of the same color.

\subsection{Gauge color code}

Being a stabilizer code, the 3D color code does not admit single-shot error correction.
Finally, we recall how a single-shot quantum error correction can be implemented in the 3D (subsystem) gauge color code. 

\subsubsection*{- Subsystem code}

A subsystem code is defined by a subgroup of Pauli operators, called the gauge group $\mathcal{G}$, that is not necessarily abelian~\cite{Poulin:2005wry}. 
The stabilizer group $\mathcal{S}$ of a subsystem code is the maximal abelian subgroup of $\mathcal{G}$ (without $-I$).
Logical qubits are encoded in the subspace that is stabilized $\mathcal{S}$.
However, only \emph{bare logical operators}, which commute with $\mathcal{S}$, but outside $\mathcal{G}$, are used to encode logical qubits~\cite{Bravyi:2011xro}. 
Consequently, bare logical operators commute with all operators in $\mathcal{G}$.

One advantage of subsystem codes is that syndrome values of stabilizer operators can be inferred by measuring gauge operators, which are typically of lower weight.
While measuring gauge operators may disturb the state, logical qubits remain unaffected because they commute with $\mathcal{S}$ and logical operators.

\subsubsection*{- Gauge color code}

The 3D gauge color code is a subsystem extension of the 3D color code that admits single-shot error 
correction~\cite{Bombin:2015tpp}. 

\begin{definition}\emph{[\textbf{3D gauge color code}]}
The gauge group $\mathcal{G}_{\text{3DGC}}$ is given by face operators with Pauli $X$ and $Z$:
\begin{eqnarray*}
\mathcal{G}_{\text{3DGC}}  &=& \langle X(f), Z(f) |f \in \Delta_2\rangle \  \\
&=&  \left\langle \figbox{0.36}{fig_2D_X.pdf}, \  \figbox{0.36}{fig_2D_Z.pdf}  \right\rangle \ .
\end{eqnarray*}
\end{definition}

The stabilizer group $\mathcal{S}_{\text{3DCC}}$ is generated by volume operators with Pauli $X$ and $Z$:
\begin{eqnarray*}
\mathcal{S}_{\text{3DCC}} &=& \langle X(c), Z(c) | c \in \Delta_3 \rangle \  \\
&=&  \left\langle \figbox{2.5}{fig_3D_X_volume.pdf}, \ \figbox{2.5}{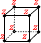}  \ \right\rangle. 
\end{eqnarray*}
Depending on the properties of the manifold where the code is defined, the stabilizer group may also contain (extended) membrane operators. 

Unlike in 2D, this gauge group $\mathcal{G}_{\text{3DGC}}$ is not abelian as some pairs of face operators may anti-commute. 
For example, two neighboring faces share a single vertex when their color labels are $(AB,CD)$, $(AC,BD)$, or $(AD,BC)$ colors. 
Then the corresponding $X(f)$ and $Z(f')$ operators anti-commute:
\begin{align}
\figbox{0.65}{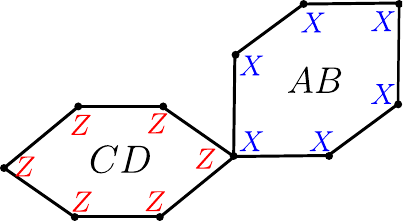}.
\end{align}

\subsubsection*{- Meta check and single-shot error correction}

Each volume stabilizer $Z(c)$ or $X(c)$ can be written as a product of face operators $Z(f)$ or $X(f)$, for example:
\begin{align}
\figbox{2.5}{fig_3D_X_volume.pdf} =
\figbox{2.5}{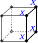} \ \times \figbox{2.5}{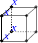} \ .
\end{align}
Therefore, only face operators need to be measured for error correction. \footnote{
Since $X(f)$ and $Z(f')$ generally do not commute, they cannot be measured simultaneously. 
Measurements must be performed in separate rounds (e.g., $X(f)$ first, then $Z(f)$). 
Using $O(1)$ rounds is allowed for single-shot error correction in the literature, and we follow this convention.
}
Moreover, volume stabilizers can be decomposed in multiple ways. For example, a volume $c_A$ with color $A$ satisfies:
\begin{align}
Z(c_A) = \prod_{f_{AB} \in c_A}Z(f_{AB}) = \prod_{f_{AB} \in c_A}Z(f_{AC}) =   \prod_{f_{AB} \in c_A}Z(f_{AD}).
\end{align}
These redundancy relations yield meta-check conditions, enabling syndrome error correction for gauge operators and robust estimation of stabilizer syndromes. It has been demonstrated that residual errors remain local~\cite{Brown_2016, Bombin:2014ksv}. 

\subsubsection*{- Logical qubits}

An important subtlety is that the number of logical qubits depends on the lattice topology. 
In fact, on closed manifolds (i.e., without boundaries), the 3D gauge color code supports no logical qubits. 
Similarly, when the 3D gauge color code is defined on a solid torus
geometry ($D_2\times S_1$) or a fattened torus geometry ($T_2\times I$), the code supports no logical qubits.
For such geometries, single-shot error correction fault-tolerantly prepares a particular codeword state of the 3D color code. 

\begin{figure}
\centering
\includegraphics[width=0.5\textwidth]{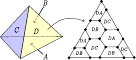}
\hspace{10pt}
\caption{A lattice with colored boundaries. Faces on the color $D$ boundary are also shown. 
}
\label{fig_lattice_boundary}
\end{figure}

A useful open manifold is shown in Fig.~\ref{fig_lattice_boundary} where each boundary is assigned a color label from ${A, B, C, D}$, such that faces on the boundary carry corresponding colors (e.g., a $D$-boundary includes $AD$, $BD$, and $CD$ faces). The lattice in Fig.~\ref{fig_lattice_boundary} has four such boundaries.
In this tetrahedral lattice, the 3D gauge color code supports one logical qubit.
The logical operators are global products over all qubits:
\begin{align}
\overline{X} = \prod_{j \in \Delta_0} X_j, \qquad \overline{Z} = \prod_{j \in \Delta_0} Z_j. 
\end{align}
The total number of qubits is always odd, ensuring that $\overline{X}$ and $\overline{Z}$ anti-commute and form a valid logical operator pair.

\subsubsection*{- Code switching}

Another important feature is that the 3D stabilizer color code can be embedded within the 3D gauge color code~\cite{Bombin:2016aoq, Anderson:2014jvy}:
\begin{align}
\mathcal{S}_{\text{3DCC}} \subset \mathcal{G}_{\text{3DGC}}.
\end{align}
That is, starting from a codeword state of the 3D gauge color code, if one measures the $Z(f)$ gauge operators and fixes the syndrome values to $+1$ via error correction, the resulting state belongs to the 3D stabilizer color code, encoding the same logical qubit.
This procedure is known as the \emph{gauge fixing}. 

In fact, the 2D stabilizer color code is also contained within the 3D gauge color code. 
Consider the following stabilizer group generated by face operators with color labels involving $D$ (i.e. $AD, BD, CD$):
\begin{align}
\mathcal{S}_{\text{2DCC}} = \langle X(f_{AD}), X(f_{BD}), X(f_{CD}), Z(f_{AD}), Z(f_{BD}), Z(f_{CD}) \rangle, \qquad f \in \Delta_2. 
\end{align}
This choice defines a 2D color code localized on the $D$ boundary, while decoupling the bulk qubits from the boundary. 
Clearly, 
\begin{align}
\mathcal{S}_{\text{2DCC}} \subset \mathcal{G}_{\text{3DGC}}
\end{align}
indicating that one can gauge-fix the 3D gauge color code to a 2D color code without altering the logical qubit.
This structure enables transitions between 3D and 2D color codes, a procedure known as \emph{code switching}, which allows the implementation of a universal gate set using transversal operations.

\section{XYZ color code without boundary}
\label{sec:XYZ}

In this section, we will mostly focus on physical and coding properties of the XYZ color code (based on qubits) when the code is supported on a closed manifold. 
%Recalling that the 3D gauge color code has no logical qubit on a closed manifold, single-shot error-correction in the XYZ code prepares only a particular codeword state fault-tolerantly.
%We will discuss the effect of boundaries in the next section.

\subsection{The model}

\subsubsection*{- XYZ color code}

\begin{definition}\emph{[\textbf{XYZ color code}]}
The XYZ color code, supported on a three-dimensional four-colorable lattice $\mathcal{L}$, is generated by face stabilizers where Pauli operators depend on the color labels as follows:
\begin{eqnarray*}
\mathcal{S}_{\mathrm{XYZ-CC}} 
&=& \langle X(f_{AB}), X(f_{CD}), iY(f_{AC}), iY(f_{BD}), Z(f_{AD}), Z(f_{BC})\rangle, \qquad f \in \Delta_2 \cr\cr 
&=& \left\langle 
\figbox{0.36}{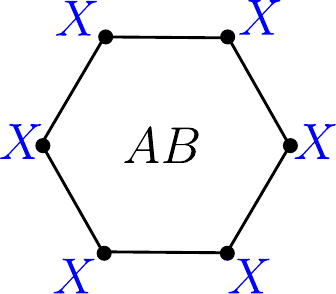},  
\figbox{0.36}{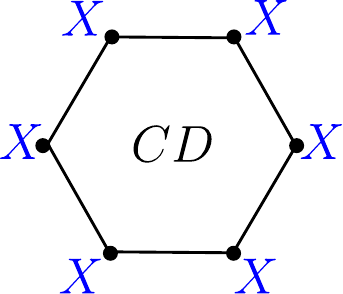}, 
\figbox{0.36}{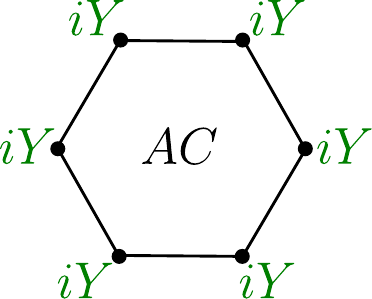}, 
\figbox{0.36}{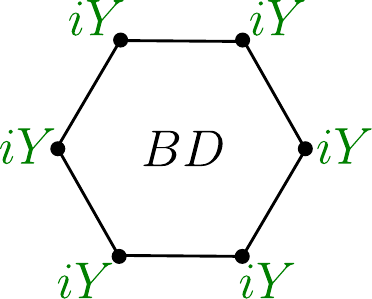}, 
\figbox{0.36}{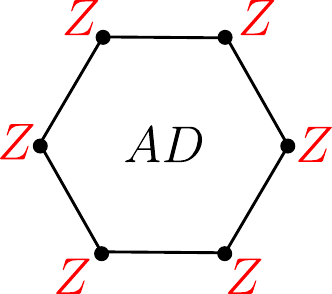}, 
\figbox{0.36}{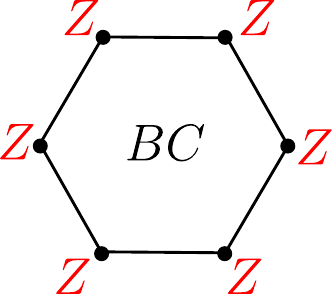} \right \rangle 
\label{eq:FTCstabilizer}
\end{eqnarray*}
\end{definition}

%This particular choice of Pauli operators ensures the commutativity of stabilizer generators as two neighboring faces share a single vertex only when their color labels are $(AB,CD)$, $(AC,BD)$, or $(AD,BC)$. 
While not explicitly included in generators, volume operators $X(c), iY(c), Z(c)$ can be found in $\mathcal{S}_{\text{XYZ-CC}}$
%\begin{align}
%X(c), iY(c), Z(c) =  \figbox{2.32}{fig_3D_X_volume.pdf}, \  \figbox{2.32}{fig_3D_Y_volume.pdf}, \ \figbox{2.32}{fig_3D_Z_volume.pdf}  \in \mathcal{S}_{\text{XYZ-CC}}, \qquad c \in \Delta_{3}
%\end{align}
as they can be expressed as products of $X(f),iY(f), Z(f)$ on each volume $c$.
Furthermore, the stabilizer group $\mathcal{S}_{\text{XYZ-CC}}$ is contained in the gauge group $\mathcal{G}_{\text{3DGC}}$ of the 3D gauge color code and thus can be gauge-fixed from the gauge color code by measuring face operators. 

\subsubsection*{- Majorana fermion code}

It is convenient to introduce a particular Majorana fermion code that can be locally mapped (gauged) to the XYZ color code.
\footnote{In fact, the 3D fermionic toric code was originally constructed this way on a cubic lattice by Levin and Wen~\cite{Levin_2003}.}
Recall that Majorana operators $\gamma_i$ obey $\gamma_i = \gamma_i^\dagger$, $\{ \gamma_i, \gamma_j  \} = 2 \delta_{ij}$ and only Majorana operators with even weights are physically admissible. 
We consider a four-colorable lattice $\mathcal{L}$ where four Majorana fermions are placed at each vertex.
Since all vertices are four-valent, one Majorana fermion can be placed at each edge with a color label of the corresponding edge ($e_{ABC},e_{ABD},\cdots$). 

\begin{definition}
\emph{[\textbf{Majorana fermion code}]}
The stabilizer group of the Majorana fermion code is given by face and vertex operators:
\begin{eqnarray*}
\mathcal{S}_{\mathrm{Majorana}}  &=& \langle S_v, S_f |v \in \Delta_0, f \in \Delta_2 \rangle \  \\
&=&  \left\langle \figbox{0.65}{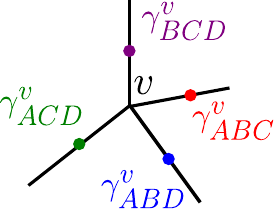}, \  \figbox{0.65}{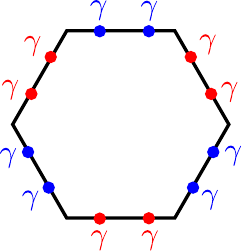}  \right\rangle \ .
\end{eqnarray*}
\end{definition}

\begin{comment}
\begin{figure}
\centering
\raisebox{\height}{\hspace{10pt}}\raisebox{-0.85\height}{\includegraphics[width=0.5\textwidth]{fig_m_lattice.pdf}}
\hspace{10pt}
\caption{The majorana fermion code. There are four majorana fermions at each vertex. 
}
\label{fig_lattice}
\end{figure}
\end{comment}

This Majorana code can be mapped to the XYZ color code by imposing a set of hard constraints on each vertex~\cite{Kitaev_2006}:
\begin{eqnarray*}
{\cal H}_{\text{fermionic}}(S_v) = \{|\psi \rangle \in {\cal H}_{\text{fermionic}} | \ \forall v \in \Delta_0: S_v |\psi\rangle = +|\psi \rangle\}.  %\longrightarrow {\cal H}_{\text{qubits}}
\end{eqnarray*}
Under this constraint, a map between two Majorana operators and a single Pauli operator can be established at each vertex (up to phases): 
\begin{align}
\figbox{0.50}{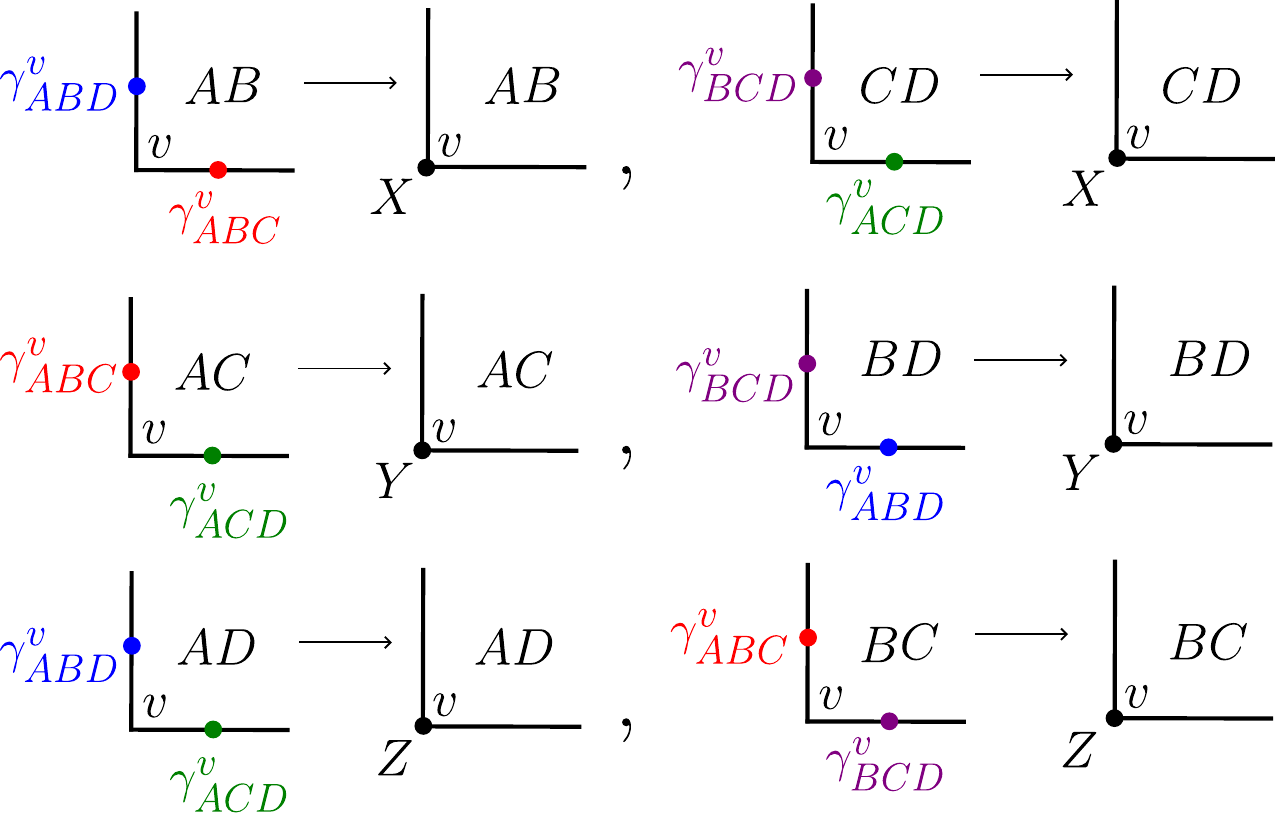} \label{eq:qubit_conversion}
\end{align}
where Majorana face operators $S_{f}$ are mapped to face operators in the XYZ-color code:
\begin{align}
S_{f_{AB}}, S_{f_{CD}}, S_{f_{AC}}, S_{f_{BD}}, S_{f_{AD}}, S_{f_{BC}}
\rightarrow
X(f_{AB}), X(f_{CD}), iY(f_{AC}), iY(f_{BD}), Z(f_{AD}), Z(f_{BC}).
\end{align}

\subsection{Fermionic excitations}\label{subsec:fermion}

\subsubsection*{- String operator}

In the XYZ color code, fermionic excitations emerge at the two end-edges of a string operator $O_{\ell}$. 
Let us begin with the Majorana fermion picture. 

\begin{definition}\emph{[\textbf{Fermion string operator}]}
Consider an open path $\ell$ that connects two distant vertices $v$ and $v'$. 
Let $e$ and $e'$ be the end-edges attached to $v$ and $v'$ respectively.
Define a string operator $O_{\ell}$ as a product of all Majorana operators along the path $\ell$:
\begin{align}
O_{\ell}\  =\  \figbox{0.70}{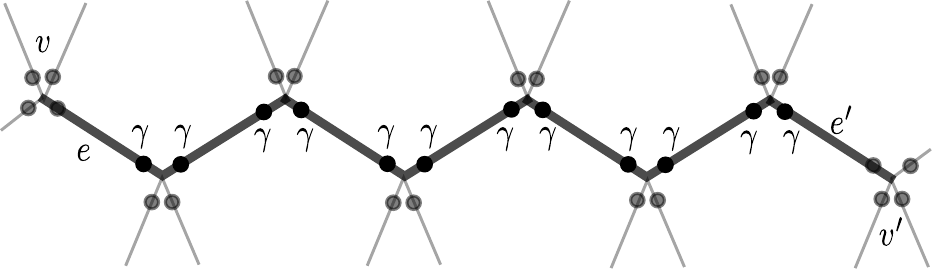} \ .
\end{align}
\end{definition}

Majorana string operator can be mapped to Pauli string operators by using the rule of Eq.~\eqref{eq:qubit_conversion}.
%, as schematically depicted below:
%\begin{align}
%O_{\ell}\ \rightarrow \figbox{0.58}{fig_qubit_string.pdf} 
%\end{align}
Note that the non-trivial support lies only on vertices between $v$ and $v'$.

\subsubsection*{- Fermionic excitations}

This string operator $O_{\ell}$ commutes with all the face operators $S_{f}$ except those contain the end-edges $e,e'$ of $\ell$.
(String-like logical operators can be constructed by considering a non-contractible loop $\ell$.)
Hence, it is natural to associate excitations to edges of the colorable lattice $\mathcal{L}$.
\footnote{There are exactly three face operators that are violated by $O_{\ell}$ at each end-edge. 
For instance, the end-edge $e_{ABC}$ is associated with three face operators $S_{f_{AB}}, S_{f_{BC}}, S_{AC}$ such that $e_{ABC}\in f_{AB}, f_{AC}, f_{BC}$. 
}

\begin{claim}\label{claim_fermion}\emph{[\textbf{Emergent fermion}]}
A string operator $O_{\ell}$, acting on a codeword state of the XYZ color code, creates a pair of fermionic excitations at end-edges $e$ and $e'$ of an open path $\ell$.
\end{claim}

The fermionic statistics can be verified by computing the statistical angle through the so-called T-junction process~\cite{Levin_2003, Kitaev_2006}. 
Let $T_{ij}$ be a hopping operator moving a particle from the position $j$ to $i$ ($O_{\ell}$ in our case). Then, the following relation arises due to the exchange statistical angle $e^{i\theta}$ of the particles:
\begin{eqnarray}
T_{43} T_{14} T_{42} = e^{i\theta} T_{42} T_{14} T_{43}
\end{eqnarray}
%Due to the deformability of hopping operators, the statistical angle $e^{i\theta}$ is an invariant under local unitary deformations~\cite{}. 
We choose the following configurations for the T-junction:
\begin{align}
\figbox{0.60}{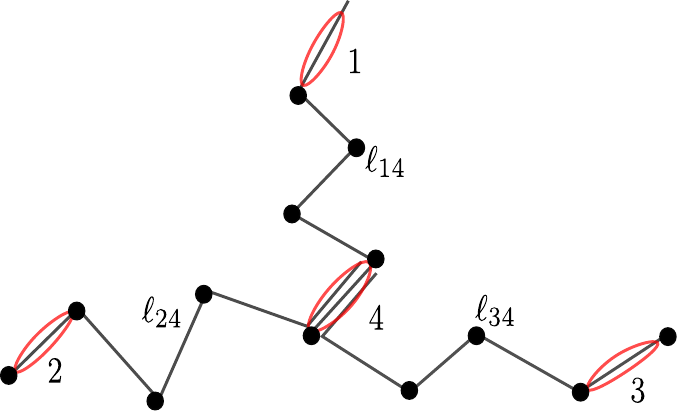}
\end{align}
where string paths $\ell_{14}, \ell_{24}, \ell_{34}$ overlap at the edge $4$. Since three hopping operators always anti-commute, an explicit calculation verifies 
\begin{align}
e^{i \theta} = \langle T_{43}^{-1} T_{14}^{-1} T_{42}^{-1} T_{43} T_{14} T_{42}\rangle  = -1 
\end{align}
confirming the fermionic statistics and claim~\ref{claim_fermion}.

\subsubsection*{- Meta checks}

The excitation configuration can be better understood by finding meta-check conditions among stabilizer generators. 
Observe that a product of face operators $S(f)$ on a single volume $c$ is an identity. 
For instance, for $c_A \in \Delta_3$, we have
\begin{eqnarray}
\prod_{f\in c_A}S(f) = \prod_{f_{AB} \in c_A } X(f_{AB}) \prod_{f_{AC} \in c_A } iY(f_{AC})
\prod_{f_{AD} \in c_A } Z(f_{AB}) = I \ .
\label{eq:local redundancy}
\end{eqnarray}
Hence, the syndrome values $s_f = 0,1$ must satisfy
\begin{align}
\sum_{f\in c} s_f = 0 \qquad (\text{mod $2$}).
\end{align}

In the dual lattice $\widetilde{\mathcal{L}}$, syndrome values $s_f$ can be associated to edges of $\widetilde{\mathcal{L}}$:
\begin{align}
\figbox{4.00}{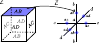}
\end{align}
where the meta-check condition becomes a vertex constraint in the $\mathbb{Z}_2$ gauge theory:
\begin{align}
\sum_{\{ \tilde{e}\ :\ v\in \tilde{e}\}} s_{\tilde{e}} = 0 \qquad (\text{mod $2$}).
\end{align}
Hence, the syndrome values $s_{\tilde{e}}$, associated to edges $\tilde{e}$ of the dual lattice $\widetilde{\mathcal{L}}$ must form closed $\mathbb{Z}_2$ loops, suggesting that all the excitations in the XYZ color code are loop-like. 

\subsubsection*{- Odd vs even loops}

%Some readers may be puzzled about claim~\ref{claim_loop} because fermionic excitations emerge as particles (zero-dimensional objects) in the Walker-Wang construction of the fermionic toric code.
%(Our model is essentially the Levin-Wen construction defined on the colorable lattice $\mathcal{L}$.)
%This apparent puzzle can be resolved by observing that 

\begin{figure}
\centering
\raisebox{\height}{a)\hspace{10pt}}\raisebox{-0.85\height}{\includegraphics[width=0.15\textwidth]{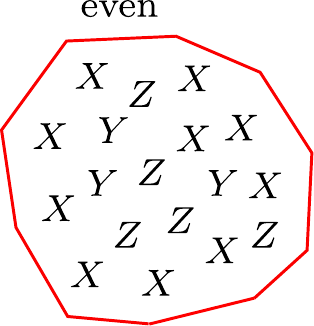}}
\hspace{10pt}
\raisebox{\height}{b)\hspace{10pt}}\raisebox{-0.85\height}{\includegraphics[width=0.4\textwidth]{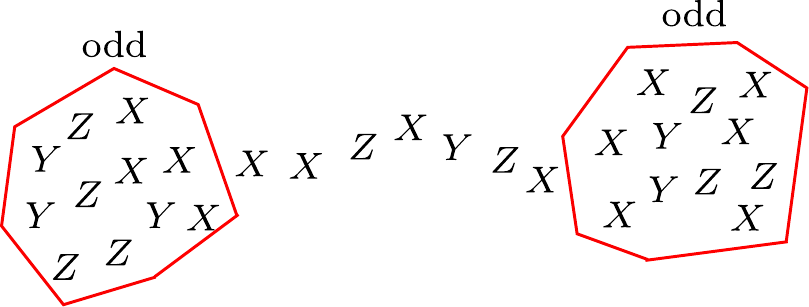}}
\hspace{10pt}
\caption{Schematic pictures of even v.s. odd loops. a) Even-loops that are created by Pauli operators acting on their interiors. 
b) Odd-loops that are connected by an open string. These loops carry fermionic statistics and cannot be annihilated locally. 
}
\label{fig_loops}
\end{figure}

There are two types of loop excitations, those with odd-weights and those with even-weights. 
First, observe that a single-body Pauli operator at each vertex creates a weight-4 loops as schematically shown below:
\begin{align}
\figbox{0.60}{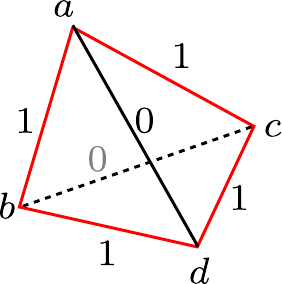}. 
\end{align}
One can enlarge this loop by attaching other weight-4 loops and create a larger closed loop with even-weights. 
Such even loops can be created by membrane-like Pauli operators, that are supported on the interior of the loops, as schematically shown in Fig.~\ref{fig_loops}(a).

On the other hand, an open string operator $O_{\ell}$ creates a pair of triangular weight-3 loops:
\begin{align}
\figbox{0.60}{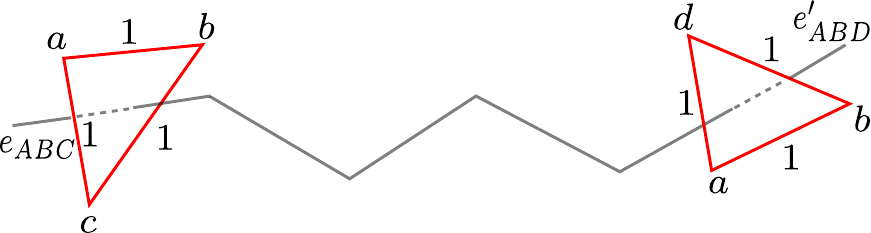}
\end{align}
In fact, one can verify that closed loops with \emph{odd-weight} always carry fermionic statistics as attaching even-weight loops to the weight-3 loop does not change the fermionic statistics. 
An important point is that these odd-weight loops cannot be created by membrane-like Pauli operators as they need to be connected by an open string operator as schematically shown in Fig.~\ref{fig_loops}(b). 
One can create a pair of weight-$3$ fermion loops by attaching bosonic weight-$4$ loops as depicted below:
\begin{align}
\figbox{0.70}{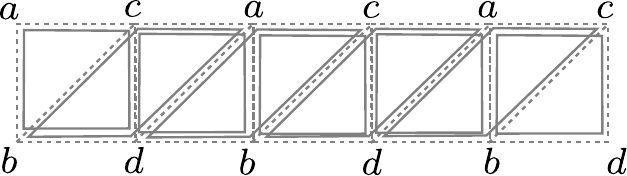}
\rightarrow 
\figbox{0.70}{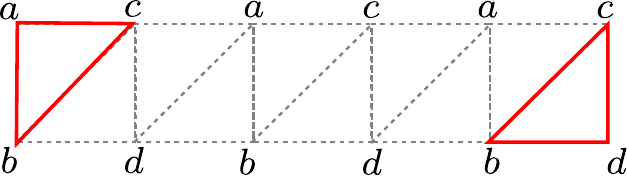}
\end{align}
Having only a single odd-loop in the whole system is prohibited due to an additional meta-check condition, which we introduce shortly.
\footnote{
One subtle point is that bosonic loops can be created by attaching fermionic excitations in the XYZ color code (as well as the Levin-Wen model). 
However, this does not appear to be the case in the Walker-Wang construction. 
Whether the XYZ color code is equivalent to the Walker-Wang construction or not (up to local unitaries) remains unclear.
A preliminary analysis suggests that translation invariance could be an obstacle for connecting two versions of the fermionic toric code.
We thank Nat Tantivasadakarn and Tyler Ellison for discussions on this point.
}

\subsection{Decoding and related issues}

\begin{figure}
\centering
\raisebox{\height}{\hspace{10pt}}\raisebox{-0.85\height}{\includegraphics[width=0.75\textwidth]{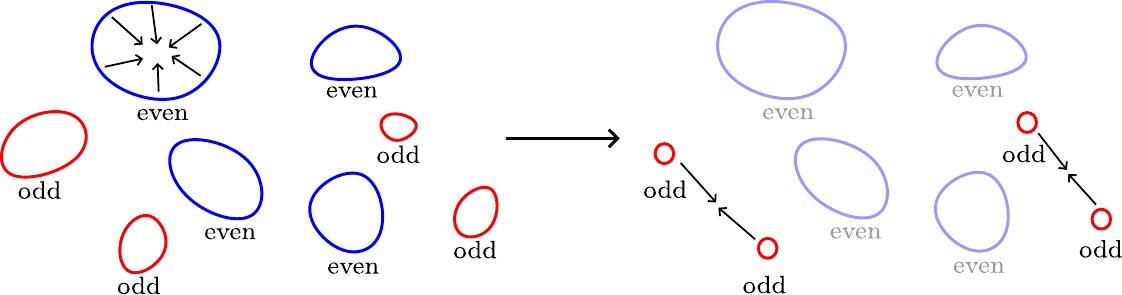}}
\hspace{10pt}
\caption{Error correction in the XYZ color code. 
}
\label{fig_decoding}
\end{figure}

We briefly discuss how to correct errors in the XYZ color code (Fig.~\ref{fig_decoding}). We will also make remarks on thermal decoherence and single-shot error correctability. 

\subsubsection*{- Stochastic noise}

Under sufficiently small local stochastic noise with perfect (i.e. no faulty) syndromes, a natural error-correction strategy is to shrink the closed loops by applying Pauli operators in their interiors. 
This eliminates all the even-weight bosonic loops, but leaves weight-$3$ fermionic loop excitations in the system. 
We then pair up fermions along the paths of minimal total weights and annihilate them. 
Since local stochastic noises are unlikely to generate large loops or widely separated fermion loops, it is expected to recover the original state with high probability. 
One can establish the presence of finite error threshold by a standard argument, bounding probabilities of large loops.
In the low error rate regime, the sizes of loops and the separation of fermions are quasi-local. 
Thus, we expect that a quasi-local quantum channel (with circuit complexity polynomial in $\log(n)$) can restore the system $\rho$ to the original state when the physical error rate is sufficiently low.

\subsubsection*{- Thermal noise}

Due to the local meta-check conditions, the XYZ color code has a finite thermal transition temperature $T_{c}$. 
At sufficiently low temperature where sizes of closed loops remain quasi-local, the thermal Gibbs state $\rho_{\beta}$ is expected to be topologically ordered (long-range entangled). 
This is because all even loops can be eliminated and all odd loops can be reduced to weight-3 fermion loops via quasi-local quantum channels. 
Since separating pairs of fermion loops do not require additional energy, fermion loops can be widely separated and thus one cannot eliminate them under local quantum channels.
The density matrix $\rho$ with low density fermion excitations, however, is still long-range entangled as the fermion statistics can be unambiguously defined by avoiding residual fermionic excitations.
Namely, the exchange statistics can be probed via the T-junction process which establishes long-range entanglement in $\rho_{\beta}$~\cite{Li:2025vru}.
One subtle point is that the XYZ color code is not a self-correcting quantum memory at finite temperature due to proliferated fermions, even though $\rho_{\beta}$ is topologically ordered. 
That is, the code only serves as a self-correcting classical memory.
This was pointed out for the fermionic toric code in a recent work~\cite{Zhou:2025bal}.

\subsubsection*{- Single-shot error correction}

When interpreted as a stabilizer code, the XYZ color code does not admit single-shot error-correction. 
The underlying reason for this is similar to the one for why the XYZ color code is not a self-correcting quantum memory. 
Namely, after correcting syndrome values via meta-check conditions, the residual errors may introduce widely separated fermionic excitations which may correspond to non-local noises. 
This problem can be avoided by realizing the XYZ color code via gauge fixing of the gauge color code. 

\subsection{Logical qubits}

\begin{claim}\label{lemma_logical_qubit}\emph{[\textbf{Logical qubits}]}
Assume that the XYZ color code is supported on a three-dimensional closed manifold. We then have
\begin{align}
k = b_2
\end{align}
where $b_2$ is the second Betti number (defined over $\mathbb{Z}_2$) which counts the number of independent non-contractible planes.
\end{claim}

On a $3$-torus, this matches with the number of logical qubits in the fermionic toric code, suggesting that the XYZ color code consists of a single copy of the fermionic toric code.
This is in contrast with the fact that the 3D color code, which consists of three copies of the 3D toric code, encodes $3b_2$ logical qubits when defined on a closed manifold.

We support this claim by explicitly counting the number of independent stabilizer generators. 
Let $V, E, F$ and $C$ be the number of vertices, edges, faces and volumes of the lattice, respectively. 
We have $V$ qubits and $F$ stabilizer generators, but not all stabilizer generators are independent. 
We need to find the number of independent stabilizer generators by identifying independent redundancy relations. 

\begin{comment}
\BY{ 
Before starting, we wish to emphasize two important subtleties in our claim. 
First, the ordinary second Betti number $b_2$ counts the number of independent non-contractible planes while $b_2$ over $\mathbb{Z}_2$ counts the number of independent mod-$2$ planes. 
While these two quantities match for a $3$-torus, they may differ in general. 
For instance, $\mathbb{R}\mathbb{P}^3$ has a vanishing Betti number, but that defined over $\mathbb{Z}_2$ is $b_2 = 1$, predicting that the XYZ color code has $k=1$ for $\mathbb{R}\mathbb{P}^3$. 
A previous work showed that the number of logical qubit in the fermionic toric code is indeed $k=1$~\cite{Barkeshli:2023bta}. 
Although this effect becomes visible when evaluating the redundancy relations on non-contractible planes (the condition 3) below), our treatment remains heuristic.
}
\end{comment}

\subsubsection*{- Derivation of $k$}

1) There are local redundancy relations associated to each volume $c \in \Delta_3$. 
Namely, for a given volume $c_A \in \Delta_3$, say with the color $A$, we have
\begin{eqnarray}
\prod_{f_{AB} \in c_A } X(f_{AB}) \prod_{f_{AC} \in c_A } iY(f_{AC})
\prod_{f_{AD} \in c_A } Z(f_{AB}) = I \ .
\label{eq:local redundancy}
\end{eqnarray}
However, not all of these local relations are independent, as we can obtain one local redundancy relation by multiplying all other local relations. Therefore, there are $C - 1$ independent local redundancy relations. 

2) There are global redundancy relations which can be obtained by multiplying all face operators sharing no colors: 
\begin{eqnarray}
 \prod_{f_{AB} \in \Delta_2} X(f_{AB})  \prod_{f_{CD} \in \Delta_2}  X(f_{CD})  =I 
\label{eq:global relation}
\end{eqnarray}
This constraint prohibits a single odd-weight loop as mentioned in section~\ref{subsec:fermion}.

Other global redundancy relations, such as those resulting from combinations $AC$ and $BD$ or $AD$ and $BC$, are not independent as they can be constructed from Eq.~\eqref{eq:global relation} together with local redundancy relations. 
Hence, there is only one independent global redundancy relation.

3) There are membrane-like redundancy relations when the lattice contains non-contractible 2D planes over $\mathbb{Z}_2$. 
Suppose $\Sigma$ be such a non-contractible plane. 
Then the corresponding redundancy relation is given by the product of all plaquette stabilizers supported on $\Sigma$:
\begin{eqnarray}
\prod_{f \in \Sigma } S(f) = I \ .
\label{eq:homological redundancy}
\end{eqnarray} 

\begin{figure}[h!]
\includegraphics[width = 15cm]{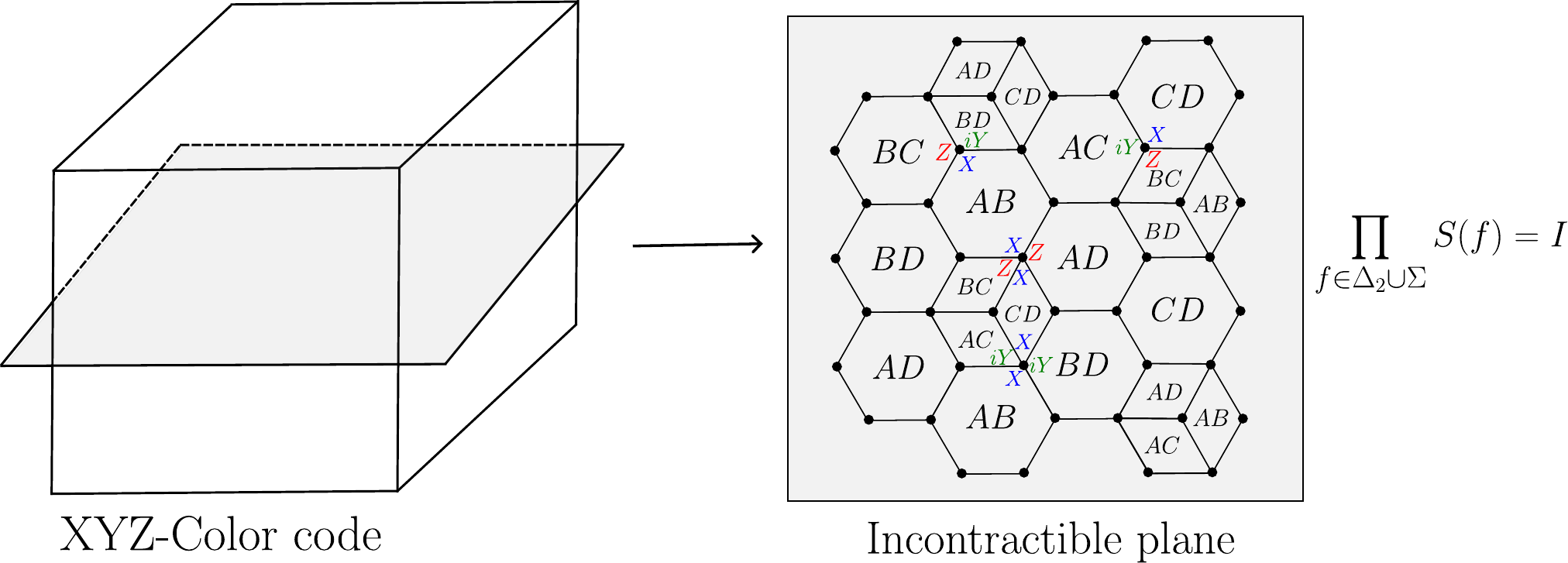}
\centering
\caption{A portion of non-contractible plane in the bulk. By multiplying all plaquette stabilizers supported on the plane, we obtain one global redundancy relation. }
\label{fig:incontractible}
\end{figure}

The condition 3) is tricky to verify. In fact, for qudit generalizations of the XYZ-color code, this condition does not necessarily hold!
To verify the redundancy condition, observe that there are two kinds of vertices on a non-contractible plane $\Sigma$ where three or four faces attached (Fig.~\ref{fig:incontractible}). 
When three faces are attached to a vertex, we always have three different Pauli operators ($X$, $iY$, and $Z$) from the face stabilizers which multiply to an identity. When four faces are attached, we always have two kinds of Pauli operators, each appearing twice (e.g. two of $X$ and two of $iY$), which multiply to an identity.
\footnote{
One potential subtlety is the overall phase factor in the redundancy relations. 
We obtain a factor of $-1$ when there are $iY$ contributions at the vertices where four faces meet. 
If the total number of such contributions turns out to be odd, the redundancy condition multiplies to $-I$. 
}  
As such, a product of face stabilizers on a non-contractible plane $\Sigma$ is an identity. 
See Fig.~\ref{fig:incontractible} for an illustration. 
Hence, there are $b_2$ redundancy relations.
\\

\begin{comment}
\BY{ 
Here, we remark on the subtlety concerning the Betti number $b_2$ by using $\mathbb{R}\mathbb{P}^3$ as an example. 
While $\mathbb{R}\mathbb{P}^3$ does not have a non-contractible plane in a conventional sense, we expect that this redundancy condition cannot be generated from local $\mathbb{Z}_2$ redundancy conditions as the second Betti number over $\mathbb{Z}_2$ is non-trivial, namely $b_2=1$ over $\mathbb{Z}_2$.  
An indirect way to see this is that $\mathbb{R}\mathbb{P}^3$ has a pair of a $\mathbb{Z}_2$ loop and a $\mathbb{Z}_2$ plane that intersect once where one can define a pair of anti-commuting logical operators. 
For this reason, we expect that the number of redundancy relations is given by $b_2$ over $\mathbb{Z}_2$. 
}
\end{comment}

Building on the above computations, the number of logical qubits encoded in this code is given by
\begin{align}
k = V - (F - C - b_2) = -V + E - F + C + b_2 = b_2, 
\end{align}
where $ -V + E - F + C =0$ for a three-dimensional closed manifold. This
leads to the claim~\ref{lemma_logical_qubit}.

\subsection{Membrane operators}

\subsubsection*{- Membrane logical operators}

Membrane-like logical operators are supported on non-contractible planes $\Sigma$.
To construct a non-trivial logical operator, we first pick a face color label and then choose a Pauli operator that differs from the native Pauli operator appearing in the face stabilizers of that color.
For instance, if we choose the color label $AB$, we must use Pauli $Z$ (or $iY$) operators, since the corresponding stabilizers are $S(f_{AB}) = X(f_{AB})$, consisting of Pauli $X$ operators.
That is, a membrane logical operator can be constructed using Pauli $Z$ (or $iY$) operators acting on all qubits associated with $AB$-colored faces lying in a non-contractible plane $\Sigma$:
\begin{align}
M_{\Sigma_{AB}}= \prod_{f_{AB} \in \Sigma } Z(f_{AB}) = \figbox{0.45}{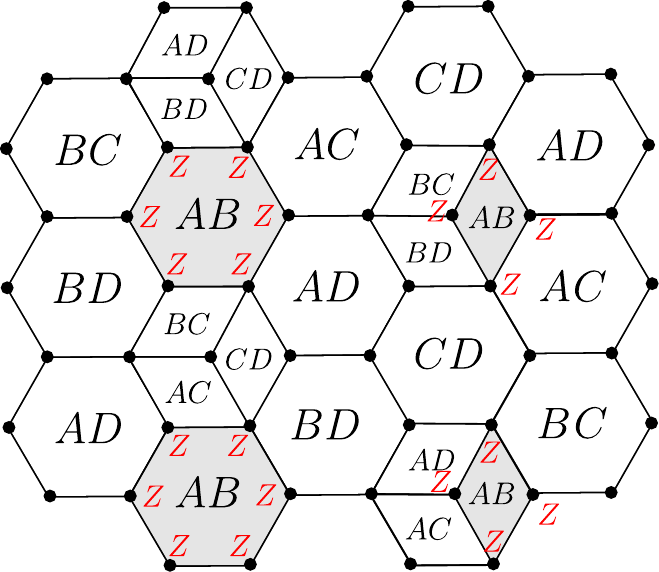}. 
\end{align}

Let us verify that $M_{\Sigma_{AB}}$ commutes with all stabilizers.
Recalling that $Z(f_{AB})$ commutes with all stabilizers except those of the form $S(f_{CD}) = X(f_{CD})$, it suffices to check that $M_{\Sigma_{AB}}$ commutes with $S(f_{CD})$.
When a face $f_{CD}$ intersects the non-contractible plane $\Sigma$, it always intersects in an even number (specifically, two) of $AB$-colored faces.
This ensures that there are always two $Z(f_{AB})$ terms in $M_{\Sigma_{AB}}$ that intersect with $X(f_{CD})$, and therefore $M_{\Sigma_{AB}}$ commutes with $S(f_{CD})$.
The non-triviality of $M_{\Sigma_{AB}}$ as a logical operator can be demonstrated by showing that it anti-commutes with a string-like logical operator $O_{\ell}$.
Membrane operators can also be constructed using different face color labels $AC$, $AD$, $BC$, $BD$, and $CD$.
However, all such membrane logical operators are equivalent to $M_{\Sigma_{AB}}$ up to multiplication by stabilizer generators.

\subsubsection*{- Open membrane operators}

The construction can be extended to any open surface $\Sigma$. 

\begin{definition}\emph{[\textbf{Membrane operator}]}
Consider a surface $\Sigma$ which consists of faces in $\Delta_2$. For each face color label, we define membrane operators by
\begin{align}
M_{\Sigma_{AB}}= \prod_{f_{AB} \in \Sigma } Z(f_{AB}), \qquad
&M_{\Sigma_{CD}}= \prod_{f_{CD} \in \Sigma } Z(f_{CD})\label{eq:AB} \\
M_{\Sigma_{AC}}= \prod_{f_{AC} \in \Sigma } X(f_{AC}), \qquad
&M_{\Sigma_{BD}}= \prod_{f_{BD} \in \Sigma } X(f_{BD}) \\
M_{\Sigma_{AD}}= \prod_{f_{AD} \in \Sigma } iY(f_{AD}), \qquad
&M_{\Sigma_{BC}}= \prod_{f_{BC} \in \Sigma } iY(f_{BC}) 
\end{align}
where in each case we choose Pauli operators that differ from those used in the native face stabilizers of the same color.
\end{definition}

\begin{figure}
\centering
\raisebox{\height}{\hspace{10pt}}\raisebox{-0.85\height}{\includegraphics[width=0.24\textwidth]{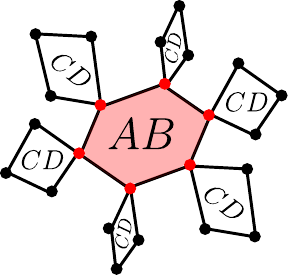}}
\hspace{10pt}
\caption{Bosonic loop excitation generated by a membrane operator $M_{\Sigma_{AB}}$. It excites stabilizers $S(f_{CD})$ located along the boundary of the membrane $\Sigma$.   
}
\label{fig_membrane_loop}
\end{figure}

These membrane operators create bosonic loops along the boundary of $\Sigma$.
For instance, $M_{\Sigma_{AB}}$ commutes with all the stabilizers except $S(f_{CD})$ lying on the boundary of $\Sigma$ as depicted in Fig.~\ref{fig_membrane_loop}. 

\subsubsection*{- Single-shot state preparation}

Finally, let us discuss the implication of single-shot error correction. 
Since the 3D gauge color code on a closed manifold has no logical qubits, single-shot error correction prepares a particular codeword state of the XYZ color code. 
One can observe that membrane logical operators of the XYZ color code are included in the stabilizer group of the 3D gauge color code on a closed manifold. 
Hence, the prepared codeword states are stabilized by membrane logical operators.

\section{XYZ color code with boundary}\label{sec:XYZ_surface}

In this section, we discuss physical and coding properties of the XYZ color code when the code is supported on an open manifold. 

\subsection{Surface fermionic topological order}

The XYZ color code realizes an anomalous surface fermionic topological order when supported on a manifold with boundary.
Consider the code defined on a solid torus, so that the boundary is a two-dimensional torus.
To be specific, we focus on the boundary labeled by color $A$, where the boundary is a hexagonal lattice, although our analysis applies to any color boundary.

\subsubsection*{- Boundary stabilizer group}

\begin{definition}\emph{[\textbf{Boundary of XYZ color code}]}
The boundary stabilizer group $\mathcal{S}^{\partial A}_{\mathrm{XYZ-CC}}$ on the $A$ boundary is generated by face stabilizers with color labels $AB,AC,AD$: 
\begin{align}
\mathcal{S}^{\partial_A}_{\mathrm{XYZ-CC}} &= \langle X(f_{AB}), iY(f_{AC}), Z(f_{AD}) \rangle  \qquad f_{AB}, f_{AC}, f_{AD} \in \partial_A \Delta_3 \\
&= \figbox{0.65}{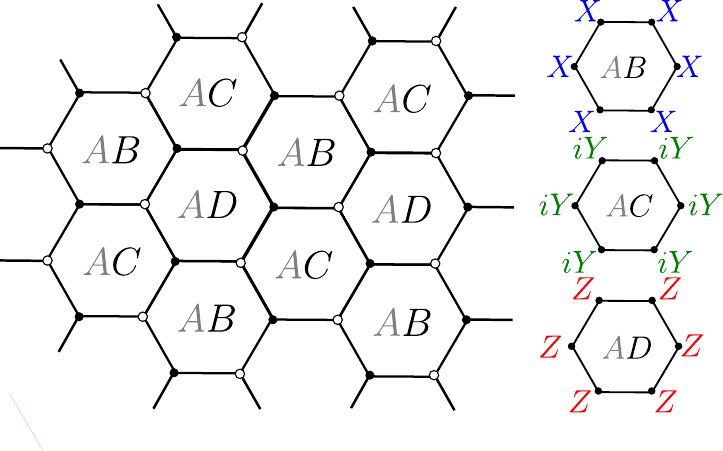} 
\end{align}
where $\partial_A \Delta_3$ denotes the 2-cells on the $A$ boundary of $\Delta_3$.
\end{definition}

This boundary stabilizer group is equivalent (up to transversal Clifford transformations) to the six-body plaquette terms of the celebrated Kitaev honeycomb model~\cite{Kitaev_2006}:
\begin{align}
S_{j} = \figbox{0.60}{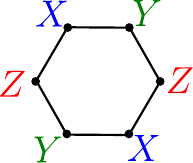} 
\end{align}
The Kitaev model itself is generated by two-body Pauli operators, while the hexagonal $S_j$ terms appear as emergent symmetries of its Hamiltonian.
One may also view this as an $XZ$-dephased variant of the $\mathbb{Z}_2$ toric code~\cite{Wang:2023uoj, Ellison_2023}.

\subsubsection*{- Long-range entangled mixed state}

When viewed purely as a stabilizer code, the boundary code $\mathcal{S}^{\partial A}_{\mathrm{XYZ-CC}}$ has small code distance $d=2$ and thus lacks intrinsic fault-tolerance. 
This follows from the fact that certain two-body operators commute with all boundary stabilizers and act as logical operators:
\begin{align}
g_{e} \ = \ \figbox{0.65}{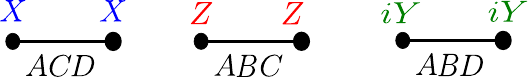} 
\end{align}
Furthermore, the boundary code encodes an extensive number of logical qubits, with $k=\frac{n_\mathrm{bdy}}{2} +1$. 
Despite this low distance and extensive degeneracy, all codeword states of the boundary code are long-range entangled~\cite{Li:2025vru}.

\begin{claim}\emph{[\textbf{Long-range entangled mixed state}]}
Consider the maximally mixed state $\rho^{\partial_A}_{\mathrm{XYZ-CC}}$ that is stabilized by the boundary stabilizer group $\mathcal{S}^{\partial A}_{\mathrm{XYZ-CC}}$:
\begin{align}
\rho^{\partial_A}_{\mathrm{XYZ-CC}} = \prod_{f \in \partial_A \Delta_3} (I + S_{f})
\end{align}
up to appropriate normalization. Then, $\rho^{\partial_A}_{\mathrm{XYZ-CC}}$ cannot be expressed as a convex sum of short-range entangled states. 
\end{claim}

This follows because $\rho^{\partial_A}_{\mathrm{XYZ-CC}}$ supports fermionic excitations whose exchange statistics can be probed by the T-junction process.
String operators that create pairs of boundary fermions can be constructed by choosing an open path on the boundary:
\begin{align}
\figbox{0.6}{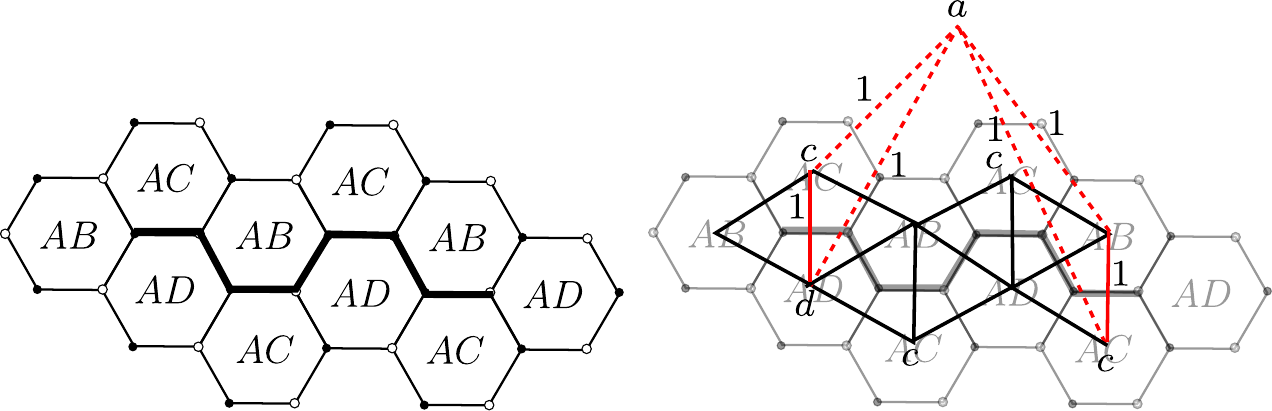}. 
\end{align}
Here, it is convenient to put a fictitious vertex with the color label $a$ to account for the boundary in the dual lattice. 
In this picture, a surface fermion corresponds to a triangular loop connecting the boundary vertex $a$ with two bulk vertices.
The non-triviality of $\rho^{\partial_A}_{\mathrm{XYZ-CC}}$ follows from the fact that the self-statistics is invariant under local unitaries~\cite{Li:2025vru}. 

\subsubsection*{- Weak vs strong symmetry}

We remark that the distinction between stabilizer and logical operators is reflected in the symmetry properties of the mixed state, namely in terms of strong and weak symmetries~\cite{Ma:2022pvq}. Stabilizer operators of the code $\mathcal{S}^{\partial A}_{\mathrm{chiral-CC}}$ act as generators of \emph{strong symmetries} of $\rho^{\partial_A}_{\mathrm{XYZ-CC}}$ since
\begin{align}
S_{f}\rho^{\partial_A}_{\mathrm{XYZ-CC}} = \rho^{\partial_A}_{\mathrm{XYZ-CC}}.
\end{align}
In contrast, two-body logical operators act as generators of \emph{weak symmetries} of $\rho^{\partial_A}_{\mathrm{XYZ-CC}}$ since
\begin{align}
g_{e}\rho^{\partial_A}_{\mathrm{XYZ-CC}} g_{e}^{\dagger}= \rho^{\partial_A}_{\mathrm{XYZ-CC}}.
\end{align}

\subsubsection*{- Single-shot preparation of fermionic mixed state}

The fact that the XYZ color code is inside the 3D gauge color code suggests that this 2D fermionic mixed state can be fault-tolerantly prepared by single-shot error correction.

\subsubsection*{- Single-shot state preparation}

Recalling that the gauge color code defined on a thickened torus has no logical qubits, single-shot error correction can prepare only a particular codeword state of the XYZ color code. 
Here we analyze which codeword state is fault-tolerantly prepared. 

Observe that the XYZ color code on a thickened torus supports $k=2$ logical qubits. 
Namely, non-contractile loop operators localized on the outer surface can be identified as logical $\overline{Z}_1$ and $\overline{Z}_2$ operators while those localized on the inner surface realize equivalent logical operators $\overline{Z}_1' \simeq \overline{Z}_1$ and $\overline{Z}_2' \simeq \overline{Z}_2$ operators.
On the other hand, membrane operators connecting two surfaces can be identified as logical $\overline{X}_1$ and $\overline{X}_2$ operators. 
Here, we find that membrane operators are inside the stabilizer group of the 3D gauge color code defined on a thickened torus.
Hence, one prepares a state stabilized by logical $\overline{X}_1$ and $\overline{X}_2$ via single-shot error correction if the code is defined on a thickened torus. 

\subsection{Majorana corners}

Finally, let us consider the XYZ color code defined on a tetrahedral lattice. In this case, the code encodes a single logical qubit, i.e. $k=1$.
This may appear puzzling at first, since the tetrahedral lattice is embedded in a 3-ball geometry (with $S^2$ boundary), on which the XYZ color code normally encodes no logical qubits.
Hence, one might think that the tetrahedral lattice is topologically equivalent to a 3-ball with trivial boundary and thus should support no non-trivial logical degrees of freedom.
However, this misses a key boundary effect that gives rise to the logical qubit.

\begin{figure}
\centering
\raisebox{\height}{\hspace{10pt}}\raisebox{-0.85\height}{\includegraphics[width=0.45\textwidth]{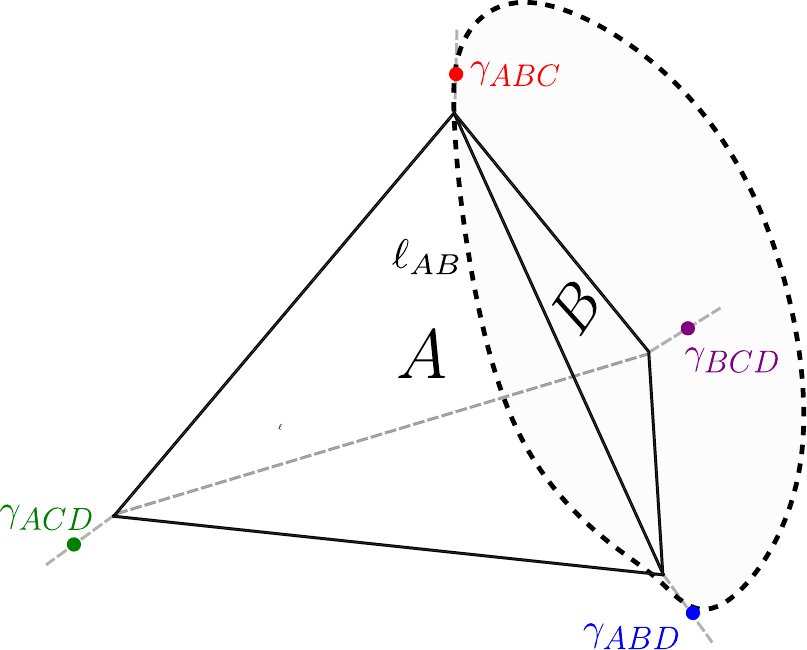}}
\hspace{10pt}
\caption{A string logical operator $O_{\ell_{AB}}$ in the XYZ color code on a tetrahedral lattice. 
This can be constructed by considering a loop $\ell_{AB}$ that connects two unpaired Majorana corner modes.
}
\label{fig_tetrahedron}
\end{figure}

On a tethrahedral lattice, the XYZ color code acquires one logical qubit in a subtle and interesting way.
As illustrated in Fig.~\ref{fig_tetrahedron}, in the Majorana fermion picture, the tetrahedral lattice features four Majorana zero modes located at the four corners.
These four dangling Majorana fermions cannot be eliminated by any local stabilizer terms and together generate a single logical qubit.
\footnote{
For the $d>2$ generalizations, $\mathbb{Z}_d$ dislocation defects~\cite{You_2012} emerge on the corners. 
}

To understand this more precisely, recall that closed string operators commute with all stabilizers.
Surface string loops that lie entirely on the boundary and do not involve the dangling Majorana modes are contractible and thus correspond to stabilizers.
However, there also exist string operators that connect the dangling Majorana modes across the corners of the tetrahedron.
These nontrivial loops define logical operators for the encoded qubit.

\begin{claim}\emph{[\textbf{Logical qubit and corner Majorana fermions}]}
Consider a path $\ell_{AB}$ connecting two Majorana corner modes $\gamma_{ABC}$ and $\gamma_{ABD}$. Then, a string operator $O_{\ell_{AB}}$ is a logical operator of the XYZ color code on a tetrahedral lattice.
Furthermore, considering string operators $O_{\ell_{CD}}, \ O_{\ell_{AC}}, \ O_{\ell_{BD}}, \ O_{\ell_{AD}}, \ O_{\ell_{BC}}$, logical Pauli operators can be constructed:
\begin{align}
\overline{X} \sim O_{\ell_{AB}}, \ O_{\ell_{CD}}, \quad
i\overline{Y} \sim O_{\ell_{AC}}, \ O_{\ell_{BD}}, \quad
\overline{Z} \sim O_{\ell_{AD}}, \ O_{\ell_{BC}}.
\end{align}
\end{claim}

\subsection{Small-size examples}

Here, we present two small-size examples of XYZ-color code. 
These examples naturally generalize to chiral color codes, which will be discussed in the next section. 

\subsubsection*{- 8-qubit code}

We first consider the code defined on a cubic lattice with eight verticies as shown below. The code has only one unit volume and three different colors are assigned at the boundary. The stabilizer group of the code is generated by plaquette operators on the boundary:
\begin{align}
\figbox{0.25}{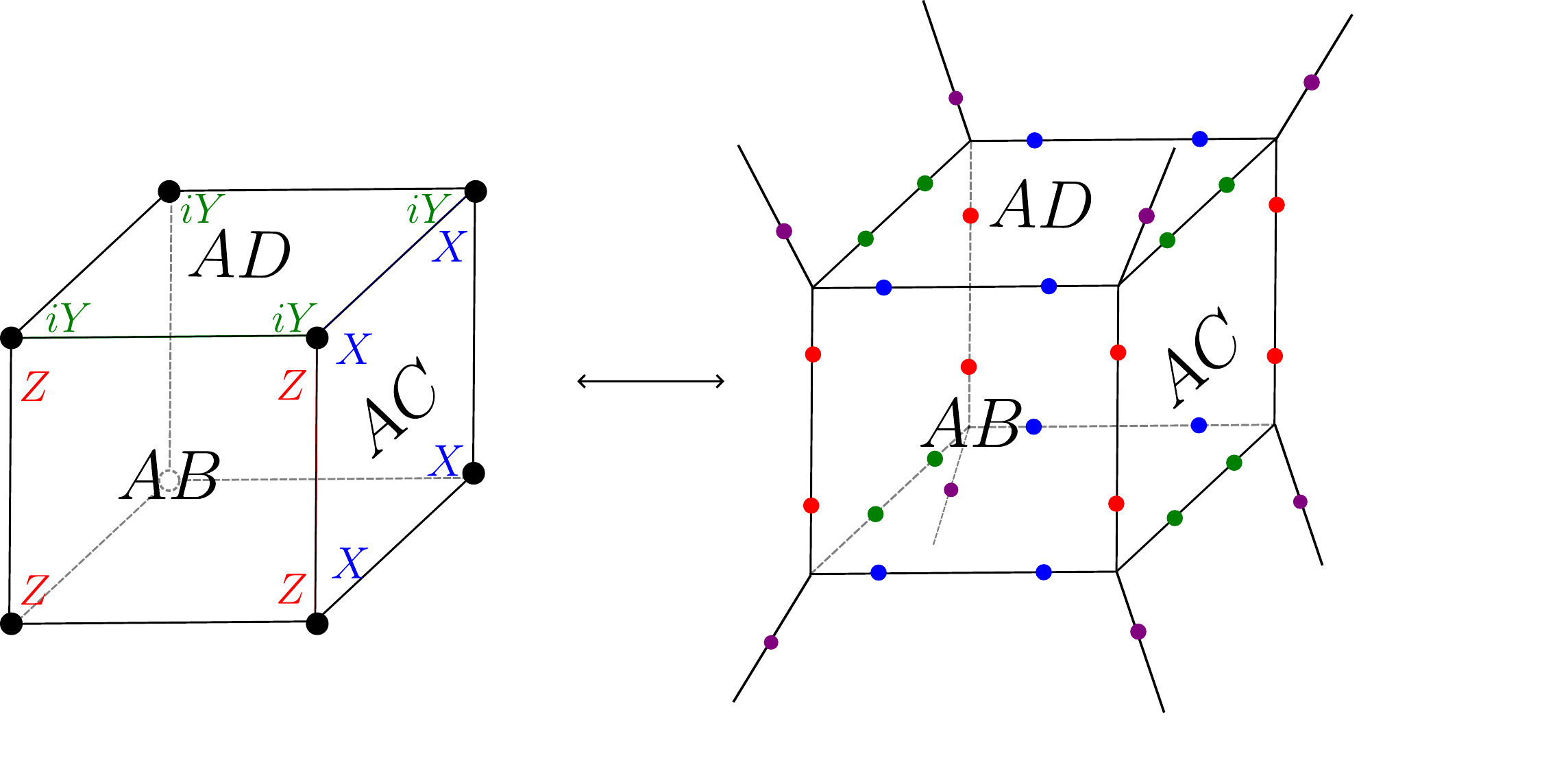}
\end{align}
The code encodes three logical qubits and the code distance is given by two, where two-body $X$ ($iY$, $Z$) operators connecting $AB$ ($AC$, $AD$) faces are logical Pauli operators. 
This code can be also viewed as Kitaev's 2D honeycomb model constructed on the 2D surface of the cube with only plaquette terms.~\cite{Kitaev_2006} 

\subsubsection*{- 15-qubit code}

Next, we consider the 15 qubit code defined on a tetrahedral lattice with fifteen vertices. There are four unit volumes with distinct colors and four color boundaries with different colors.  
\begin{align}
\figbox{0.5}{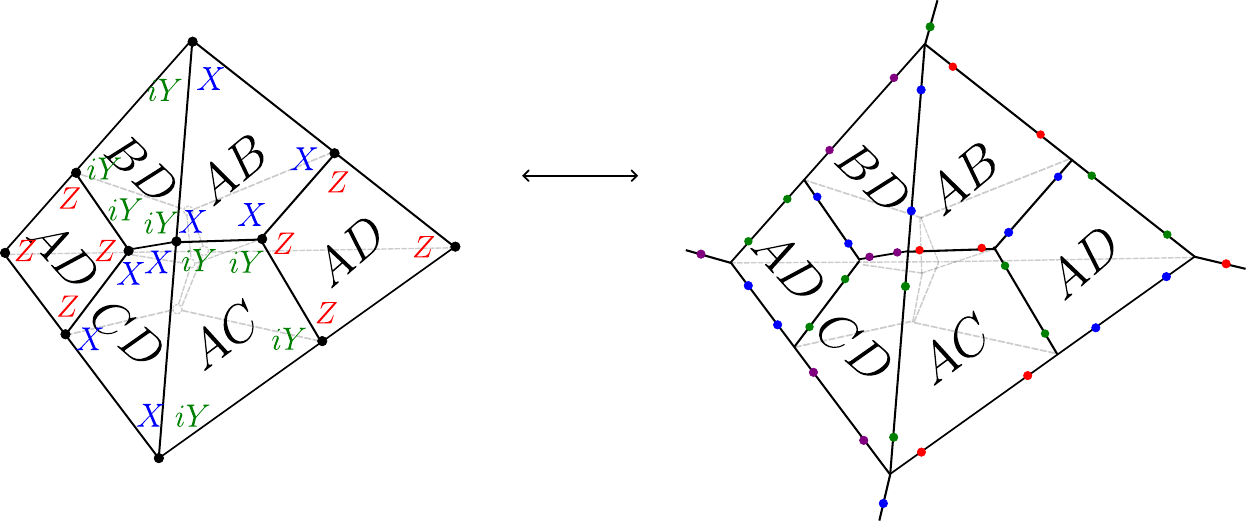}
\end{align}
The code encodes one logical qubit with Pauli logical operators $\bar{X} = \prod_{v \in \Delta_0} X_v$ and $\bar{Z} = \prod_{v \in \Delta_0} Z_v$. The code distance is given by three, where the smallest logical operator is given by a weight-three fermionic string operator connecting two corners. 

%These logical operators exhibit fermionic statstics when we perform T-junction process as in the section~\ref{sec:3C}. Since they are logical operators, fermions emerge inside the code space rather than excitations in this case.\footnote{This code can be also viewed as Kitaev's 2D honeycomb model constructed on the 2D surface of the cube with only plaquette terms.~\cite{Kitaev_2006}} 

%\section{Hybrid code: fermion-to-qubit conversion }

%Given that single-shot code-switching between the 2D color code and the 3D color code already achieves transversal implementation of universal logical gates, we naturally wonder if the XYZ color code provides any additional advantage in fault-tolerant quantum information processing. 
%Here, we propose that the XYZ color code may work as an efficient and fault-tolerant converter of quantum information between qubit and fermionic systems. 
%Given that qubit and fermionic architectures have their own strengths, fault-tolerant conversions between the two would be potentially beneficial. 
%Also, for fermionic systems, not much is known about transversal logical gates. 

%One concrete realization of our proposal is a hybrid quantum code, which is defined on a system consisting both of qubits and fermions. 
%One useful example of such a code can be constructed by recalling the mapping between the majorana fermion code and the XYZ code. 
%Here

%\addcontentsline{toc}{section}{Part II : Qudit generalization}
%\section*{Part II : Qudit generalization}

\section{Chiral color code without boundary}\label{sec:chiral}

In this section, we present a qudit extension of the XYZ color code, called the chiral color code, and discuss its physical and coding properties. 
We will focus on the cases when the code is supported on a closed manifold. 

\subsection{The model}

\subsubsection*{- Colorable lattice is bipartite}

To construct the qudit chiral color code, it is crucial to recall that colorable lattices are bipartite~\cite{kubica2018abcs}. 
Namely, one can assign a bipartition label $\lambda(v)=\pm 1$ to each vertex $v$ such that neighboring vertices always have opposite signs.
We will use black and white dots in the figures to represent the $\pm 1$ assignments:
\begin{align}
\figbox{0.5}{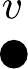}\ \rightarrow \ \lambda(v)= +1 \qquad \figbox{0.5}{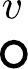}\ \rightarrow \lambda(v) =  -1. 
\end{align}

For a given $i$-cell $\delta \in \Delta_i$, we define the action of a Pauli operators $X^a Z^b$ on all qubits on vertices of $\delta$ as
\begin{align}
X(\delta) \equiv \prod_{v \in \delta} X_v, \qquad 
Z(\delta) \equiv \prod_{v \in \delta} Z_v^{\lambda(v)}
\end{align}
where $X,Z$ are qudit generalizations of Pauli operators with $ZX = \omega XZ$, $\omega = e^{2\pi i/d}$, and $d$ is the qudit dimension. 
In this convention, $X(\delta)$ acts uniformly as $X$ on all vertices in $\delta$, while $Z(\delta)$ acts as $Z$ or $Z^{-1}$ with a sign determined by the bipartition label $\lambda(v)$ at each vertex:
\begin{align}
X(f) = \figbox{0.35}{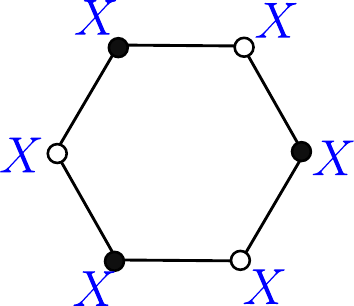}, \qquad Z(f) = \figbox{0.35}{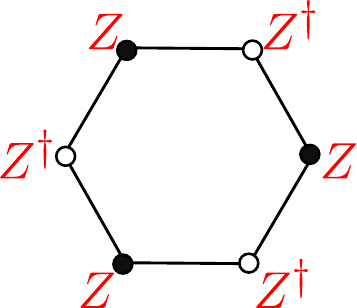} \ .
\end{align}
With these operators, the qudit 2D and 3D color codes can be defined in a way similar to the qubit cases.
See~\cite{Watson2015} for discussions on coding properties of the qudit color code.

\subsubsection*{- Chiral color code}

\begin{definition}
\emph{[\textbf{Chiral color code}]}
The $\mathbb{Z}_d^{(\alpha)}$ chiral color code, supported on a three-dimensional four-colorable lattice $\mathcal{L}$ with $d$-dimensional qudits, is generated by face stabilizers where qudit Pauli operators depend on the color labels as follows:
\begin{eqnarray*}
&&\mathcal{S}_{\mathrm{chiral-CC}} 
= \langle X(f_{AB}), X(f_{CD}), 
X^{-1}Z^{-\alpha}(f_{AC}), X^{-1}Z^{-\alpha}(f_{BD}), 
Z^{\alpha}(f_{AD}), Z^{\alpha}(f_{BC})\rangle
, \qquad f \in \Delta_2 \cr\cr 
&=& \left\langle \figbox{0.33}{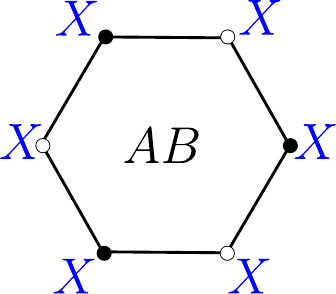},  
\figbox{0.33}{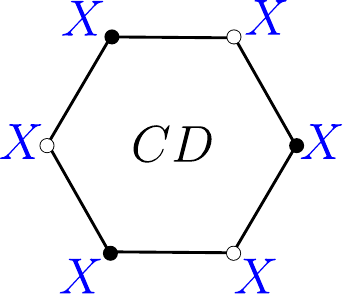}, \figbox{0.33}{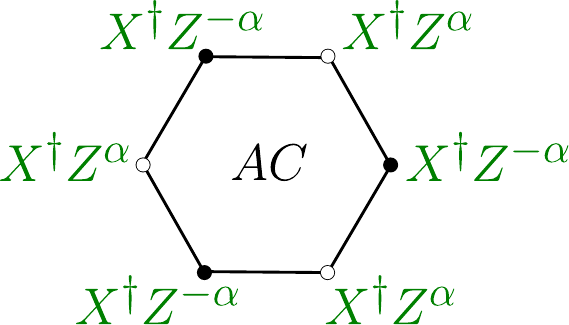}, \figbox{0.33}{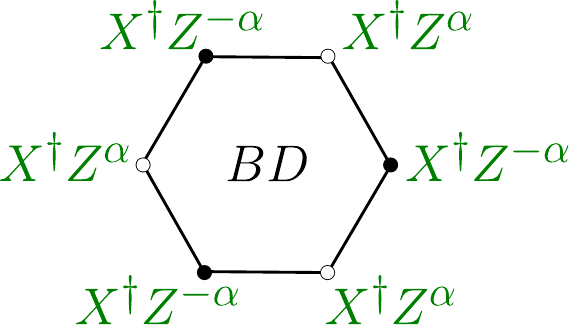}, \figbox{0.33}{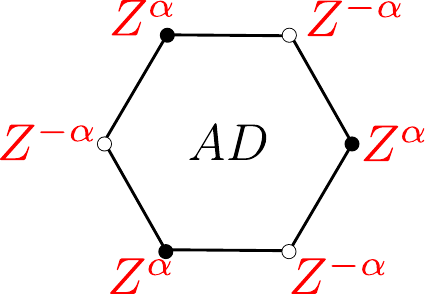}, \figbox{0.33}{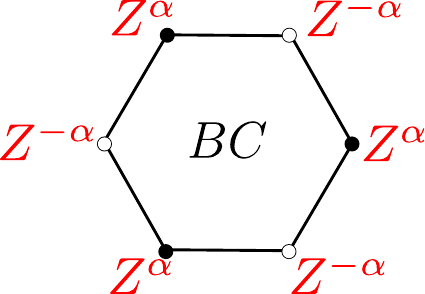}  \right \rangle
\label{eq:qudit model}
\end{eqnarray*}
\end{definition}

This particular choice of Pauli operators, with the bipartition labels, ensure the commutativity of stabilizer generators. 
Volume operators $X(c), \ X^{-1}Z^{-\alpha}(c), \ Z^{\alpha}(c)$ can be found inside the stabilizer group $\mathcal{S}_{\mathrm{chiral-CC}}$, while the stabilizer group $\mathcal{S}_{\mathrm{chiral-CC}}$ is contained inside the gauge group $\mathcal{G}_{\mathrm{3DGC}}$ of the 3D qudit gauge color code.
\footnote{ 
The chiral color code features topological order only when $d$ and $\alpha$ are coprime, as otherwise $AD$ and $BC$ plaquette stabilizers do not have order $d$. However, the construction can be simply generalized for the case when $d$ and $\alpha$ are non-coprime, by dropping $\alpha$ exponent on the $AD$ and $BC$ plaquette stabilizers. 
}

\subsubsection*{- Overview of physical and coding properties}

The chiral color code exhibits a rich variety of physical and coding properties, depending on the choice of $d$ and the chirality parameter $\alpha$.  

\begin{claim}\emph{[\textbf{Chiral anyons}]}\label{claim:WW}
The $\mathbb{Z}_d^{(\alpha)}$ chiral color code realizes the 3D topological order of the Walker-Wang model based on the modular category of $\mathbb{Z}_d^{(\alpha)}$ anyons.
\end{claim}

See~\cite{Bonderson:2007zz} for detailed study of the $\mathbb{Z}_d^{(\alpha)}$ anyons. 
We will support this claim by explicitly verifying key characteristics of the $\mathbb{Z}_d^{(\alpha)}$ Walker-Wang model in the rest of the paper. 

\begin{itemize}
\item The code supports a chiral surface topological order with the $\mathbb{Z}_d^{(\alpha)}$ anyons.
\item For odd $d$, the bulk hosts no point-like excitations. When defined on a closed manifold, the code encodes no logical qudits. 
\item For odd $d$, there exists a local quantum channel that prepares the ground state on a closed manifold. That is, the bulk is short-range entangled.
\item For $d \equiv 2 \ (\mathrm{mod}\ 4)$, the bulk supports the fermionic toric code. On a closed manifold, the code encodes $k = b_2$ logical qubits (not qudits).
\item For $d \equiv 0 \ (\mathrm{mod}\ 4)$, the bulk supports the bosonic toric code, also with $k = b_2$ logical qubits on a closed manifold.
\end{itemize}

Given the simplicity of our construction, it is remarkable that the chiral color code faithfully captures the rich physics of the $\mathbb{Z}_d^{(\alpha)}$ Walker–Wang model. In fact, it is expected that all 3D topological orders built from modular categories of abelian anyons can be realized by stacking multiple copies of $\mathbb{Z}_d^{(\alpha)}$ theories and applying boson condensation, as seen from the Witt group consideration~\cite{Haah:2019fqd}. 
Consequently, many of the fault-tolerance features we identify in this work can be viewed as universal properties of 3D abelian topological phases, including those with chiral anyons. This is especially valuable from both a physical and quantum information viewpoint. 
Furthermore, the original Walker–Wang construction is extremely complex~\cite{Walker:2012mcd}, and prior to our model, its stabilizer-like realizations remained elusive except specific instances.

\subsection{Excitations}

The local redundancy (meta-check) conditions in the chiral color code are similar to those in the XYZ color code.
In particular, the product of all face stabilizers around a closed volume $c$ equals the identity:
\begin{eqnarray}
\prod_{f\in c_A}S(f) = I.
\end{eqnarray}
The syndrome values $s_f$ (or equivalently $s_{\tilde{e}}$ assigned to edges in the dual lattice $\widetilde{\mathcal{L}}$) obey the vertex constraint:
\begin{align}
\figbox{4.60}{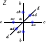}\qquad \sum_{\{ \tilde{e}\ :\ v\in \tilde{e}\}} s_{\tilde{e}} = 0 \qquad (\text{mod $d$}).
\end{align}

While this vertex constraint may appear similar to the vertex constraint in the $\mathbb{Z}_d$ gauge theory, there is a crucial difference. 
\footnote{
In the gauge theory, edges are oriented and the vertex constraint includes $\pm$ signs based on those orientations. 
In contrast, the chiral color code possesses the bipartition labels $\lambda(v)=\pm 1$ while it lacks the orientation structure.
}
The constraints in the chiral color code reflects the bipartite lattice structure and allows only loops with alternating $+1$ and $-1$ syndrome patterns:
\begin{align}
\figbox{4.0}{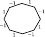}.
\end{align}
Crucially, these loops are admissible only if their total length is even.

One important implication is that triangular loops are not always allowed when $d > 2$.
To see this, consider the following ansatz for a triangle:
\begin{align}
\figbox{4.60}{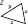}\ , \qquad x + y = y + z = z + x = 0 \qquad  (\text{mod $d$}).
\end{align}
This suggests $y = -x$, $z = -x$, and $y = -z$, hence 
\begin{align}
2x = 0 \qquad \text{mod $d$}.
\end{align}

\textbf{Case 1: $d$ even.}
The equation $2x = 0$ admits nontrivial solutions:
\begin{align}
\figbox{4.60}{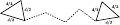}
\end{align}
These triangular excitations behave like $\mathbb{Z}_2$-charged particles and must appear as pairs at the endpoints of string operators.
Depending on the value of $d$, they may carry either bosonic or fermionic statistics, as we will explore further.

\textbf{Case 2: $d$ odd.}
The constraint $2x = 0$ forces $x = 0$, i.e., the trivial solution only.
Thus, triangular excitations are forbidden in the bulk for odd $d$, implying the absence of point-like excitations.
Instead, string-like excitations emerge, where the endpoints are not particles but rather nonlocal triangular loops with a connecting string:
\begin{align}
\figbox{4.60}{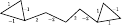}
\end{align}
When the system has a boundary, however, the interior portion of the string may condense at the boundary. In such cases, triangular loops at the ends can exist as point-like excitation on the surface and exhibit chiral anyonic statistics.

\begin{claim}\emph{[\textbf{Bulk particle}]}
The chiral color code supports non-trivial point-like excitations in the bulk if and only if $d$ is even. 
\end{claim}

\subsection{String and membrane operators}

String-like excitations in the chiral color code can be created by appropriately defined string operators. 
We begin by introducing the notion of \emph{native Pauli operators}, denoted by $P(f)_v$.

\begin{definition}\emph{[\textbf{Native Pauli operator}]}
Let $v$ be a vertex and $f$ be a face containing $v$.
We define a native Pauli operator $P(f)_v$ as a single-qudit Pauli operator acting on the vertex $v$, where the specific Pauli depends on the color label of $f$ and the bipartition sign $\lambda(v)=\pm 1$:
\begin{align}
\figbox{0.45}{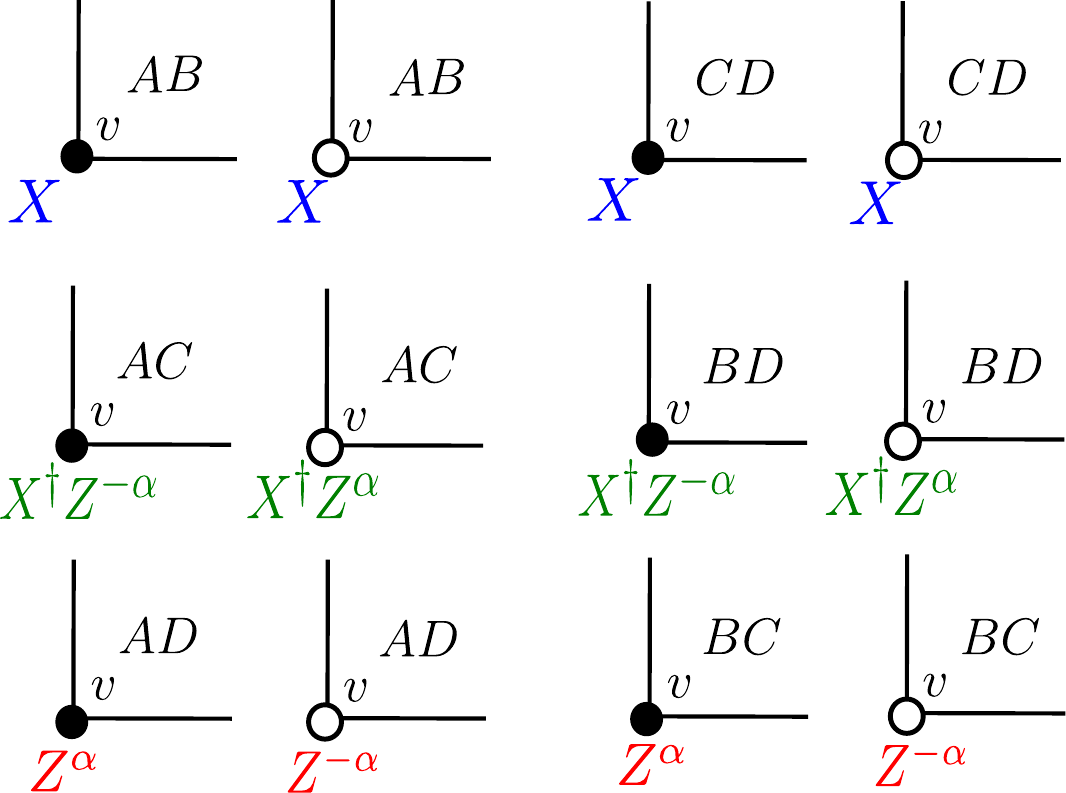}
\end{align}
\end{definition}

Native Pauli operators correspond to those appearing in face stabilizers. 
Given native Pauli operators, string operators are constructed as follows: 

\begin{definition}\emph{[\textbf{String operator}]}
Consider an open path $\ell$ that connect two distant vertices $v_0$ and $v_0'$. Define the string operator $O_{\ell}$ as a product of native Pauli operators along all intermediate vertices on $\ell$:
\begin{align}
O_{\ell} = \prod_{v \in \ell / v_0,v_0'}P(f[\ell])_v = \figbox{0.5}{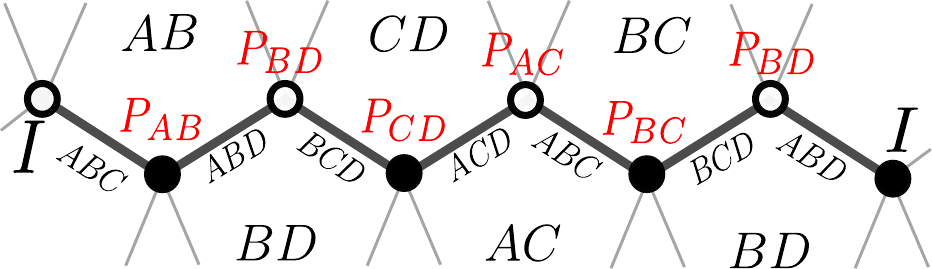} \label{eq:bulk_string}
\end{align}
where the face color label $f[\ell]$ is determined by the color labels of the two adjacent edges at each vertex $v$.
\end{definition}

If $\ell$ is a closed path encircling a single face $f$, then $O_\ell$ corresponds to the stabilizer generator:
\begin{align}
S(f_{AB}) = X(f_{AB}) = \prod_{v \in f_{AB}} P(f_{AB})_v. 
\end{align}
However, unlike the $d = 2$ case, $O_\ell$ generally creates excitations when $\ell$ traverses multiple faces.
Specifically, for an open path $\ell$, the operator $O_\ell$ creates string-like excitations with a factor of $\alpha$:
\begin{align}
O_{\ell} \rightarrow  \figbox{4.5}{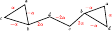} \ .
\end{align}
The face color labels of the resulting triangular loop excitations depend on the end edges of $\ell$. 

When $d$ is even, string operators of the form ${O_{\ell}}^{\frac{d}{2}}$ generate point-like $\mathbb{Z}_2$ excitations, as the intermediate portion of the string operator cancels out. For these excitations, we compute the exchange statistics explicitly using the T-junction method. We find that the particles exhibit bosonic statistics when $d \equiv 0 \pmod{4}$ and fermionic statistics when $d \equiv 2 \pmod{4}$.
\\

Membrane operators for the chiral color code resemble those for the $d=2$ cases with one important subtlety. 
Recall that membrane operators can be constructed from local meta-check (redundancy) conditions. 
For instance, volume stabilizer operators with the colors $A$ and $C$ multiplies to an identity:
\begin{align}
\prod_{c_{A}\in \Delta_3} X(c_A)
\prod_{c_{C}\in \Delta_3} X^{-1}(c_C) = I
\end{align}
where the use of $X$ and $X^{-1}$ for $A$ and $C$ volumes respectively is crucial as this ensures that their product cancel with each other. 
The product taken over a connected region $R$, consisting of $A$ and $C$-volumes, then creates Pauli-$X$ membrane stabilizer operators on the surface $\Sigma = \partial R$ where Pauli $X$ or $X^{-1}$ operators are supported on faces $f_{AC}$ with color $AC$: 
\begin{align}
M_{\Sigma_{AC}}= \prod_{f_{AC} \in \Sigma } X^{ \gamma(f_{AC}) }(f_{AC}), \qquad \gamma(f_{AC}) = \pm 1.
\end{align}
Namely, the exponent $\gamma(f_{AC}) = \pm 1$ depends on whether $A$ volume, adjacent to the face $f_{AB}$, is placed inside or outside of the surface $\Sigma$.

\begin{definition}\emph{[\textbf{Membrane operator}]}
Consider a surface $\Sigma$ which consists of faces in $\Delta_2$. For each face color label, we define membrane operators by
\begin{align}
M_{\Sigma_{AB}}= \prod_{f_{AB} \in \Sigma } Z^{\alpha \gamma(f_{AB})}(f_{AB}), \qquad
&M_{\Sigma_{CD}}= \prod_{f_{CD} \in \Sigma } Z^{\alpha \gamma(f_{CD})}(f_{CD}) \\
M_{\Sigma_{AC}}= \prod_{f_{AC} \in \Sigma } X^{ \gamma(f_{AC})}(f_{AC}), \qquad
&M_{\Sigma_{BD}}= \prod_{f_{BD} \in \Sigma } X^{\gamma(f_{BD})}(f_{BD}) \\
M_{\Sigma_{AD}}= \prod_{f_{AD} \in \Sigma } X^{-\gamma(f_{AD})}Z^{-\alpha\gamma(f_{AD})}(f_{AD}), \qquad
&M_{\Sigma_{BC}}= \prod_{f_{BC} \in \Sigma } X^{-\gamma(f_{BC})}Z^{-\alpha \gamma(f_{BC})}(f_{BC})\ 
\end{align}
where $\gamma(f)=\pm1$ depends on which volume color label is inside (or outside) of $\Sigma$. 
For each membrane operator, the associated Pauli operator differs from the corresponding face stabilizer.
\end{definition}

One can verify that these membrane operators create closed loop excitations:
\begin{align}
M_{\Sigma_{AB}} \rightarrow \figbox{4.5}{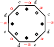}\ .
\end{align}
where the signs $\mp 1$ of excitations depend on the bipartition sign $\lambda(v)=\pm 1$ of the vertex $v$ adjacent to the perimeter of the membrane $\Sigma$.

A particularly interesting feature arises for odd $d$. 
Closed loop excitations can also be created by appropriately powered closed string operators. 
Specifically, when $\ell$ is a closed loop, we find:
\begin{align}
O_{\ell}^{\frac{d+1}{2}} \rightarrow \figbox{4.5}{fig_chiral_loop_alpha.pdf}. \label{eq:local_loop}
\end{align}
Thus, $O_{\ell}^{\frac{d+1}{2}}$ creates the same excitation as the membrane operator $M_{\Sigma_{AB}}$ create. 
This observation will be particularly useful when proving that the chiral color code is short-range entangled.

\subsection{Logical qubit/qudit}

While the construction might appear to be a natural generalization of the qubit XYZ model, its coding properties differ dramatically when the code is defined on a closed manifold. 
In fact, the code supports logical \emph{qubit} (not qudit!) only when $d$ is even.

\begin{claim}\emph{[\textbf{Logical qubits}]}
Assume that the $\mathbb{Z}_d^{(\alpha)}$ chiral color code is supported on a three-dimensional orientable closed manifold. Then, we have 
\begin{align}
k_{\mathrm{qubit}} &= b_2   \qquad (d = \mathrm{even}) \\
&= 0 \qquad (d = \mathrm{odd})
\end{align}
where $k_{\mathrm{qubit}}$ is the number of logical qubit.
\end{claim}

This result can be derived similarly to the qubit cases by counting the number of independent redundancy relations. 
However, there are subtle differences between the qubit cases and the odd $d$ cases. 
Recall that, for the qubit cases, there are local redundancy relations associated to each volume $c \in \Delta_3$ which
leads to local syndrome constraints:
\begin{align}
\sum_{f \in c} s_f = 0 \qquad (\mbox{mod $2$}).
\end{align}
But not all of these local constraints are independent as combining all the local syndrome constraint leads to a trivial relation:
\begin{align}
\sum_c \sum_{f \in c} s_f  = 0 \qquad (\mbox{mod $2$})
\end{align}
where the LHS is \emph{identically zero} as $s_f$ appears twice in the sum. 
Hence, this equation does not constrain possible values of $s_f$.
This suggests there are only $C - 1$ local redundancy relations. 
In addition, there is one global redundancy relation. Hence, in total, there are $C$ independent local and global redundancy relations for the qubit cases. 

On the contrary, when $d$ is odd, all the local redundancy relations become independent. 
This can be seen by observing that combining local syndrome constraints does not give trivial relation. 
One can verify that the following equation has a trivial solution only:
\begin{align}
\sum_c g(c)\sum_{f\in c} s_f =0 \qquad (\mbox{mod $d$})
\end{align}
where $g(c)$ is mod $d$ variables.
Namely, this equation can be satisfied as an identity only when $g(c)=0$ for all $c$, suggesting that there are $C$ local redundancy relations. 
But in this case, there is no independent global redundancy relation as a global redundancy relation can be generated by combining local redundancy relations. Hence, in total, there are $C$ independent local and global redundancy relations when $d$ is odd. 

The case where $d$ is even (but not two) can be treated as a hybrid of the above two situations. 
One can verify that there are $C$ independent local and global redundancy relations.

When the code is defined on a homologically nontrivial manifold, key difference is also in the membrane-like redundancy relations.
For odd $d$, membrane-like redundancy does not exist. 
To see this, consider the product of face operators on a non-contractible surface $\Sigma$:
\begin{align}
\prod_{f \in \Sigma} S(f) \not=I.
\end{align}
While Pauli operators cancel at every trivalent vertex of $\Sigma$, they fail to cancel at four-valent vertices.
As a result, the only redundancy available is of order two:
\begin{align}
\prod_f S(f)^{\frac{d}{2}} =I \qquad \text{for even $d$}.
\end{align}
This gives rise to a $\mathbb{Z}_2$-valued logical qubit per homologically nontrivial plane.
Thus, the number of encoded logical qubits is $k_{\mathrm{qubit}} = b_2$ for even $d$, and zero otherwise.

\section{Bulk short-range entanglement}\label{sec:chiral_SRE}

In this section, we show that the chiral color code for odd $d$ is short-range entangled when defined on a closed manifold by constructing explicit local quantum channel preparing its unique ground state $|\psi\rangle$. 

\subsection{Preparation by local quantum channel}

\begin{definition}\emph{[\textbf{Short-range entanglement}]}
A pure state wavefunction $|\psi\rangle$ is said to be short-range entangled if there exists a local quantum channel $\mathcal{Q}$ such that
\begin{align}
\mathcal{Q}\left(|0\rangle\langle 0|^{\otimes n}\right) = |\psi\rangle\langle\psi|.
\end{align}
That is, there exists a local unitary $U$ acting on the system and ancillary qudits such that
\begin{align}
\mathrm{Tr}_a\left[U\left(|0\rangle\langle 0|^{\otimes n}_s \otimes |0\rangle\langle 0|^{\otimes n_a}_a\right)U^\dagger\right] = |\psi\rangle\langle\psi|.
\end{align}
\end{definition}

Conventionally, a pure state $|\psi\rangle$ is regarded as short-range entangled if it can be prepared from a product state by a finite-depth local unitary circuit. 
Here, we extended this notion by allowing preparation via a local quantum channel, i.e., by including the use of ancillary qudits together with local unitaries and partial trace.

\begin{claim}\emph{[\textbf{Trivial bulk}]}
Let $d$ be an odd integer. When defined on a closed three-dimensional manifold, the $\mathbb{Z}_d^{(\alpha)}$ chiral color code (coprime $d$ and $\alpha$) has a unique ground state $|\psi\rangle$ that is short-range entangled. 
\end{claim}

The use of local quantum channels, rather than unitaries, is essential here. The existence of a local unitary that prepares $|\psi\rangle$ from a product state without ancillas remains open.~\footnote{We note that, although our channel prepares the bulk ground state, it does not prepare the full bulk Hamiltonian. This can be verified by observing that excited states cannot be generated by flipping zeros in the initial product state.} We expect that no finite-depth unitary can prepare the $\mathbb{Z}_d^{(\alpha)}$ chiral color code Hamiltonian from a trivial Hamiltonian, as that would imply the existence of a two-dimensional commuting-projector Hamiltonian realizing the $\mathbb{Z}_d^{(\alpha)}$ topological order, which is believed to be impossible in general due to the absence of gapped boundaries or Lagrangian subgroups in these theories.
\begin{comment}
For most choices of $(d,\alpha)$, we expect that no finite-depth local unitary circuit can prepare $|\psi\rangle$ from a product state.
Such a local unitary would imply the existence of a two-dimensional commuting-projector Hamiltonian realizing the $\mathbb{Z}_d^{(\alpha)}$ topological order, which is believed to be impossible in general due to the absence of gapped boundaries or Lagrangian subgroups in these theories.
\end{comment}
\footnote{
A conjecture along this line is generally supported by an observation that string-net type constructions always have the Lagrangian subgroup and thus gappable boundaries. An alternative way is to show the incompatibility of the nonzero Hall conductance and commuting-projector construction (see~\cite{Hall} for instance).
However, a quantum state can be ``chiral'' without possessing nonzero Hall conductance, see~\cite{Vardhan:2025oyx} as well as our upcoming work.
}
\footnote{An important exception is the $\mathbb{Z}_{p^{2m}}^{(1)}$ theory for odd prime $p$, which has a Lagrangian subgroup and hence admits a gapped boundary and a commuting-projector Hamiltonian in two dimensions, despite that the anyon set can be chiral in a sense that the anyon content is not invariant under time reversal. This will be discussed in an upcoming work.}

\subsubsection*{- Comment on QCAs}

Previous results on short-range entanglement in the Walker–Wang models relied on constructions using quantum cellular automata (QCAs), which belongs to a certain subset of local quantum channels and act as locality-preserving operator mapping.
A non-trivial QCA is not implementable via a local unitary circuit, but can be implemented via a local quantum channel by including trivial ancilla systems.  
Known examples include the 3-fermion model~\cite{Haah:2018jdf}, the chiral semion model~\cite{Shirley:2022lhu}, and the $\mathbb{Z}_p^{(1)}$ Walker–Wang model for prime $p$~\cite{Haah:2019fqd}. (See also~\cite{Bauer_2023})
\footnote{
We thank Nat Tantivasadakarn and Yu-An Chen for explaining their upcoming work on QCAs. 
}
These models generate a wide variety of modular abelian topological orders under boson condensation and stacking, forming a Witt class~\cite{Haah:2019fqd}. 

While the QCA approach is particularly useful when the system possesses translation symmetries, a direct and general argument of short-range entanglement for arbitrary $\mathbb{Z}_d^{(\alpha)}$ topological orders, beyond a few limited cases with translation symmetries, has been lacking. 
Our work provides the explicit and fully general construction of a local quantum channel that prepares the ground state for all odd $d$ and arbitrary $\alpha$ (coprime to $d$).
This not only establishes the bulk short-range entanglement of the $\mathbb{Z}_d^{(\alpha)}$ chiral color codes on closed manifolds, but also demonstrates that such exotic phases admit fault-tolerant quantum information processing.
We believe this is the most comprehensive result on short-range entanglement in Walker-Wang-type models.
One limitation, however, is that our construction does not directly reduces to QCA constructions. 

More broadly, multiple copies of the $\mathbb{Z}_d^{(\alpha)}$ model can give rise to an even richer class of modular abelian anyon theories.
For example, the chiral semion theory can be obtained by condensing a boson in a single copy of $\mathbb{Z}_4^{(1)}$, while the 3-fermion model arises from a specific boson condensation pattern involving four copies of $\mathbb{Z}_4^{(1)}$, see~\cite{Ellison_2023} for instance.
Our result applies uniformly to such composite constructions too. 
The only essential condition is the absence of transparent bosons or fermions (bulk anyons that braid trivially with all others) since their presence would imply residual bulk topological order and obstruct short-range entanglement.

Hence, the long-standing belief, concerning bulk short-range entanglement in the Walker-Wang-type model, can be shown in a unified footing via explicit construction of local quantum channels. 
Furthermore, on a practical end, local quantum error-correction procedures for preparing $|\psi\rangle$ can be performed fault-tolerantly via single-shot error correction within the framework of the 3D gauge color code. 

\subsection{Local channel from local error-correction}

We begin by presenting a quantum error-correction procedure for preparing the ground state $|\psi\rangle$ of the chiral color code on a closed manifold. 
Starting from the product initial state $|0\rangle^n$, we measure all face stabilizers $S(f)$ to obtain a syndrome configuration $\vec{s}=\{ s_f \}$ and then apply a recovery operation $V(\vec{s})$ to eliminate excitations and restore the ground state:
\begin{align}
|0\rangle^n \underset{\text{measure}\ S(f)}{\longrightarrow} |\psi(\vec{s})\rangle \underset{\text{feedback} \ V(\vec{s})}{\longrightarrow} |\psi(0)\rangle.
\end{align}
In general, this procedure does not yield a local quantum channel because the feedback operation $V(\vec{s})$ may depend on global features of the syndrome configuration $\vec{s}$. 
For instance, this is the case in preparing the ground state of the 2D toric code, where the correction must pair up all excitations that may be far separated from each other.

To obtain a genuinely local quantum channel, the correction $V(\vec{s})$ must be computable as a local $O(1)$-body function of the syndrome data $\vec{s}$. That is, the correction at each site should depend only on nearby syndrome outcomes within a finite neighborhood.

This error-correction procedure can be implemented unitarily using ancilla qudits and trace operations. Specifically, we begin with the initial state $|0\rangle^n \otimes |0\rangle^{n_a}$ in the system and ancilla qudits where system qudits live on the vertices of the lattice $\mathcal{L}$, and one ancilla qudit is associated to each face. 
We then record the stabilizer measurement outcomes in the ancilla by performing controlled-$X$ operations conditioned on the eigenvalues of $S(f)$.
This procedure can be implemented by local Clifford unitary acting jointly on the system and ancillas:
\begin{align}
|0\rangle^n \otimes |0\rangle^{n_a} \underset{\text{local Clifford}}{\longrightarrow}  \sum_{\vec{s}} |\psi(\vec{s})\rangle \otimes |\vec{s}\rangle. 
\end{align}
Here $\psi(\vec{s})$ is the post-measurement state with syndrome $\vec{s}$:
\begin{align}
|\psi(\vec{s})\rangle \propto \prod_f \pi_f(s_f) |0\rangle^n, \qquad 
\pi_f(s_f) = \frac{1}{d}\sum_j \omega^{-s_f j }S(f)^j
\end{align}
where $\pi_f(s_f)$ is a projector onto the $\omega^{s_f}$ eigenspace of $S(f)$.
Note that only syndrome values $\vec{s}$ consistent with the meta-check constraints contribute to the sum, and all admissible $\vec{s}$ appear with equal amplitude due to the structure of the Clifford circuit and the stabilizer state.

The remaining step is to find a local feedback operator $V(\vec{s})$ that transforms $|\psi(\vec{s})\rangle$ to $|\psi\rangle$ 
\begin{align}
V(\vec{s})|\psi(\vec{s})\rangle  = e^{i\theta(\vec{s})}|\psi\rangle
\end{align}
up to a phase factor $e^{i\theta(\vec{s})}$.
In the next subsection, we demonstrate that such a $V(\vec{s})$ can be constructed, depending only on $\vec{s}$ in a finite neighborhood. 
Applying this correction unitarily on the system, we obtain:
\begin{align}
 \sum_{\vec{s}} |\psi(\vec{s})\rangle \otimes |\vec{s}\rangle \underset{\text{local Clifford}}{\longrightarrow} 
|\psi\rangle \otimes  \sum_{\vec{s}}  e^{i\theta(\vec{s})} |\vec{s}\rangle,
\end{align}
which completes the preparation of $|\psi\rangle$ via a local unitary circuit with ancilla and partial trace.

\subsection{Removing loops locally}

Recall that face stabilizers are given by 
\begin{align}
&S(f_{AB}) = X(f_{AB}), \qquad 
S(f_{AC}) = X^{-1}Z^{-\alpha}(f_{AC}), \qquad
S(f_{AD}) = Z^{\alpha}(f_{AD}) \\  
&S(f_{CD}) = X(f_{CD}), \qquad  S(f_{BD}) = X^{-1}Z^{-\alpha}(f_{BD}), \qquad  S(f_{BC}) = Z^{\alpha}(f_{BC}).
\end{align}
For the initial state $|0\rangle^n$, $S(f_{AD})$ and $S(f_{BC})$ already stabilize the state. 
We now present a sequence of local procedures that eliminates all excitations except those associated with $S(f_{CD})$. \\

\textbf{1) Elimination of $S(f_{AC})$ excitations:} Observe that a single-qudit $Z$ operator, applied at a vertex, creates a square-loop excitation:
\begin{align}
\figbox{4.5}{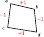}.
\end{align}
This property allows us to locally remove all $S(f_{AC})$ excitations without disturbing face stabilizers $S(f_{AD})$ and $S(f_{BC})$, which are already corrected. 
Importantly, this correction depends only on the local syndrome value at each $S(f_{AC})$ face and can be applied independently across the system.

\textbf{2) Elimination of $S(f_{AB})$ excitations on each $A$-colored volume:}
After correcting $S(f_{AC})$, the only potentially violated stabilizers on each $A$-colored volume are the $S(f_{AB})$ terms. 
(recall that $S(f_{AD})$ excitations are already corrected) 
However, the product of all face stabilizers on an $A$ volume satisfies a redundancy condition:
\begin{align}
\prod_{f \in c_A} S(f) = I. 
\end{align}
This implies that the total syndrome values of $S(f_{AB})$ on each $A$-color volume must sum to zero modulo $d$.
Crucially, these excitations can be locally removed using two-qudit $Z$ operators $Z(e_{ACD})$ (with the bipartition sign $\lambda(v)=\pm$ dependence) acting within each $A$-volume.
Since these operators commute with already-corrected stabilizers, the correction is both local and non-disruptive. 

At this stage, the excitations of $S(f_{AB})$, $S(f_{AC})$, $S(f_{AD})$, and $S(f_{BC})$ are corrected, and only $S(f_{BD})$ and $S(f_{CD})$ may remain violated.

\textbf{3) Elimination of $S(f_{BD})$ excitations on each $B$-colored volume:}
A similar argument applies to the $B$-colored volumes. Once $S(f_{AB})$ and $S(f_{BC})$ are corrected on a $B$-volume, any residual $S(f_{BD})$ excitation can be locally removed by applying a $Z$ operator on the corresponding edge $e_{ABC}$. This step does not affect previously corrected stabilizers.
\\

After the steps above, all stabilizers are corrected except possibly those of type $S(f_{CD})$. 
These remaining excitations are supported on the $CD$-colored faces of the lattice. 
It will be convenient to analyze the problem in the dual lattice $\widetilde{\mathcal{L}}$ where stabilizer violations of $S(f_{CD})$ correspond to excitations living on dual edges of color label $cd$.
Namely, the sublattice $\widetilde{\mathcal{L}}_{cd}$, formed by $c,d$ dual vertices, is a bipartite graph, so all closed paths have even length.
The syndrome redundancy relations in this sublattice are equivalent to those of a conventional $\mathbb{Z}_d$ gauge theory.
For instance, by assigning an orientation (e.g., $c \leftarrow d$), we can consistently treat excitations as $\mathbb{Z}_d$ charges and identify loops.

Hence, our task is reduced to locally removing $\mathbb{Z}_d$-charged loops in $\mathcal{L}_{cd}$.  
Note that this is not possible in the conventional $\mathbb{Z}_d$ toric code since eliminating a loop requires applying Pauli operators inside the loop.
Namely, identifying the interior region of a closed loop requires global features of the syndrome configurations. 
On contrary, in the case of the chiral color code, one can eliminate a loop by applying a closed string operator along the loop, not in the interior, as in Eq.~\eqref{eq:local_loop}. 
By using this property, all the $\mathbb{Z}_d$-charged loops can be removed locally. 

\begin{figure}
\centering
\raisebox{\height}{a)\hspace{5pt}}\raisebox{-0.85\height}{\includegraphics[width=0.45\textwidth]{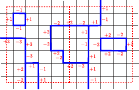}}
\hspace{10pt}
\raisebox{\height}{b)\hspace{5pt}}\raisebox{-0.85\height}{\includegraphics[width=0.45\textwidth]{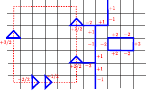}}
\hspace{10pt}
\caption{a) Example of $\mathbb{Z}_d$ loops and cubic decomposition (depicted in 2D for clarity). 
b) Application of string operators $O_{\ell}$ within a block eliminates internal loops and leaves triangular charges at the interface (e.g, $+3/2$, $+2/2$ modulo $d$). 
These triangular charges at the interface can be cancelled by similar procedures applied to adjacent cubes. 
}
\label{fig_cubic_partition}
\end{figure}

To demonstrate this explicitly, we partition the system into sufficiently large but finite cubic blocks, as illustrated in Fig.~\ref{fig_cubic_partition}(a).  
Here, loops entirely contained within a block can be eliminated locally.
For loops that pass between blocks, we apply string operators $O_{\ell}$ along the paths:
\begin{align}
\figbox{4.5}{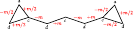}
\end{align}
with appropriate value of $m$ where $\pm m/2$ should be understood modulo $d$. 
Such operators eliminate the loops but leave behind triangular charges at the interfaces between two blocks as illustrated in Fig.~\ref{fig_cubic_partition}(b). 
By applying similar string operators in adjacent blocks, these triangular charges at the interfaces can be exactly cancelled by choosing the path $\ell$ appropriately. 
In this way, removals of local loops within blocks eliminate all $S(f_{CD})$ excitations. 

It is crucial that loop excitations can be removed by string operators. 
If only membrane-like operators could create or annihilate loops, then eliminating a large loop would require shrinking it step by step which requires global syndrome information. 
The use of string operators ensures that loop removal remains fully local and parallelizable.\\

We conclude with a remark on the residual ancilla state $|\phi\rangle \equiv \sum_{\vec{s}} e^{i\theta(\vec{s})} |\vec{s}\rangle$ produced in the course of the preparation circuit. 
Given that the system state $|\psi\rangle$ realizes the $\mathbb{Z}_d^{(\alpha)}$ topological order, it is natural to conjecture that the ancilla state $|\phi\rangle$ encodes the complex conjugate topological order, namely the $\mathbb{Z}_d^{(-\alpha)}$ topological order.

To verify this conjecture, the phase factor $e^{i\theta(\vec{s})}$ will play an important role. 
While the local error-correction procedure can returns $|\psi(\vec{s})\rangle$ to the ground state $|\psi\rangle$ up to a phase, it cannot detect global features such as \emph{twists} in loop-like excitations. 
These undetectable twists lead to nontrivial phases in the correction operator, resulting in the mismatch of $e^{i\theta(\vec{s})}.$
Thus, the structure of $e^{i\theta(\vec{s})}$ will encode precisely the information about loop twists and modular data that characterizes $\mathbb{Z}_d^{(-\alpha)}$. 
Namely, the exchange statistics of underlying chiral anyons could be directly determined from $e^{i\theta(\vec{s})}$.
A further analysis of $|\phi\rangle$ may illuminate this duality and clarify the role of ancillas for preparing chiral topological order.

\section{Chiral color code with boundary}\label{sec:chiral_surface}

In this section, we demonstrate that the boundary of the chiral color code supports anomalous $\mathbb{Z}_d^{(\alpha)}$ anyons. 

\subsection{$\mathbb{Z}_d^{(\alpha)}$ anyons}

The $\mathbb{Z}_d^{(\alpha)}$ anyons arise as a particular subset of the $\mathbb{Z}_d$ toric code anyons. 
Letting $e$ and $m$ be charge and flux of the $\mathbb{Z}_d$ toric code, the set of toric code anyons is given by
\begin{align}
\mathcal{A}_{d}^{\mathrm{toric}} \equiv \{ e^{i}m^j  \}_{i,j=0,\cdots, d-1}.
\end{align}
The $\mathbb{Z}_d^{(\alpha)}$ anyon theory is generated by the composite anyon:
\begin{align}
a \equiv e m^{\alpha}
\end{align}
and its anyon content is 
\begin{align}
\mathcal{A}_{d}^{(\alpha)} \equiv \{ a^i \}_{i=0,\cdots, d-1}\subset \mathcal{A}_{d}^{\mathrm{toric}}.
\end{align}
The topological spin and braiding statistics of abelian anyons are given by
\begin{align}
\theta(a^i) = \omega^{\alpha i^2}, \qquad B(a^i, a^j) = \omega^{2 \alpha i j }\label{eq:braiding} 
\end{align}
where $\theta(g)$ is the self-statistics and $B(g,h)$ is the mutual braiding phase between $g$ and $h$. The chiral central charge can be computed using topological spins of anyons~\cite{Kitaev_2006}.  
\begin{eqnarray}
e^{\frac{2\pi i}{8} c_- } = \frac{1}{\sqrt{d}} \sum_{i = 0}^{d-1} \theta(a^i) 
= \frac{1}{\sqrt{d}} \sum_{i = 0}^{d-1} \omega^{\alpha i^2} 
\end{eqnarray}
The quadratic Gauss sum has a closed form expression that depends on $d$ and $\alpha$. When $d = 1$ (mod 4), $c_- = 0$ (mod 8) if $\alpha$ is a quadratic residue modular $d$, otherwise, $c_- = 4$ (mod 8). When $d = 3$ (mod 4), $c_- = 2$ (mod 8) if $\alpha$ is a quadratic residue, otherwise $c_- = 6$ (mod 8). When $d$ is even, we can similarly compute the chiral central charge after condensing transparent bosons ($d = 0$ (mod 4)) or fermions ($d=2$ (mod 4)).
 
\begin{claim}
When the chiral color code is defined on a manifold with boundaries, it supports anomalous surface topological order with the $\mathbb{Z}_d^{(\alpha)}$ anyon.
\end{claim}

Note that $d=2$ and $\alpha=1$ corresponds to the fermionic toric code.

\subsection{Surface chiral topological order}

We now verify that the chiral color code supports the $\mathbb{Z}_d^{(\alpha)}$ anyons. 
We begin by constructing string operators on the boundary that create chiral surface anyons and characterize their topological properties.

Consider the $A$-colored boundary. Observe that the product of all face stabilizers along the boundary satisfies
\begin{align}
\prod_{f \in \text{$A$-boundary}} S(f) = I.
\end{align}
This suggests that one can form a strong symmetry loop stabilizer operator by multiplying $S_f$ on a connected region $R$:
\begin{align}
\hat{O}_{\ell} = \prod_{f\in R} S_f
\end{align}
where $\ell$ represents the boundary (loop) of $R$. 
Truncating this stabilizer operator $\hat{O}_{\ell}$ creates open string operators:
\begin{align}
\hat{O}_{\ell} = \figbox{3.0}{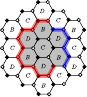} \ .
\end{align}
Here, we used simplified notations $B,C,D$ to denote face color labels $AB,AC,AD$. 

By construction, this surface string operator $\hat{O}_{\ell}$ commutes with all face stabilizers and edge stabilizers (those extending into the bulk) except those at the endpoints.
It is useful to explicitly find excitation patterns created by a string operator:
\begin{align}
\figbox{3.0}{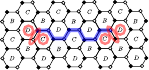}\ 
\end{align}
where violations of stabilizers are marked by circles.
The expectation values of violated stabilizers are given by
\begin{align}
\figbox{3.0}{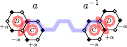}
\end{align}
where the exponent $j$ in $\omega^j$ are shown (so they are $\omega^{\alpha}$ or $\omega^{-\alpha}$). They can be identified as chiral anyons $a$ and $a^{-1}$ in $\mathbb{Z}_{d}^{(\alpha)}$ as we will explicitly verify below.
In the dual lattice picture, the excitation configuration is given by
\begin{align}
\figbox{0.6}{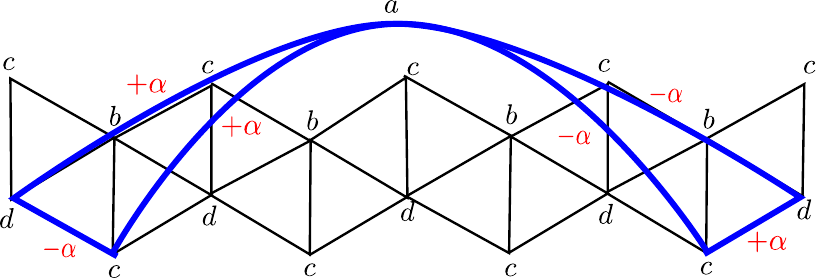} \ 
\end{align}
which lie at the endpoints of the string in the dual lattice. 
The vertex $a$ in this figure corresponds to the $A$-colored boundary.
Here, it is crucial to observe that these excitations exist because of the boundary, as they satisfy the meta-check constraints only in the presence of the dual boundary vertex $a$ and cannot be realized in a closed manifold.

\begin{figure}
\centering
\raisebox{\height}{\hspace{5pt}}\raisebox{-0.85\height}{\includegraphics[width=0.9\textwidth]{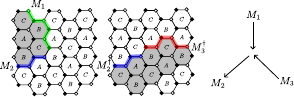}}
\hspace{10pt}
\caption{
Construction of string operators $M_1, M_2, M_3$ for T-junction.
}
\label{fig_T_junction_alpha}
\end{figure}

Anyon charges can be characterized by the braiding statistics which can be explicitly evaluated by using these surface string operators. 
Let $S_{\ell}$ be a loop stabilizer which encloses an anyon $a$:
\begin{align}
\figbox{3.0}{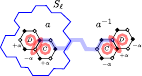}.
\end{align}
Recalling that $W_{\ell}$ is a product of $S_f$ inside $\ell$, we have
\begin{align}
B(a,a) = \omega^{2\alpha}. 
\end{align}
Topological spin (the exchange statistics) of an anyon can be probed by the T-junction process by considering three string operators $M_1, M_2, M_3$ (Fig.~\ref{fig_T_junction_alpha}):
\begin{align}  
\theta(a) = \Tr\Big[ M_3^{\dagger} M_2^{\dagger} M_1^{\dagger}  M_3 M_2 M_1 \rho_{\mathcal{S}}\Big] = \omega^{\alpha}.
\end{align}
The simplest way to verify this is to perform the T-junction process for two-body surface string operators $XX$, $Z^{\alpha}Z^{- \alpha}$, and $X^{-1}Z^{-\alpha}X^{-1}Z^{\alpha}$.
We can also verify $\theta(a^j) = \omega^{\alpha j^2}$.
Note that the angle $\theta(a)$ is a topological invariant and does not depend on details of how one chooses $M_1, M_2, M_3$. An example of construction of $M_1, M_2, M_3$, emerging from particular loop stabilizer operators, is shown in Fig.~\ref{fig_T_junction_alpha}.
This confirms that the surface anyons in the chiral color code is given by the $\mathbb{Z}_d^{(\alpha)}$ anyon theory. 

A subtle, but important point is that the surface string operator $\hat{O}_{\ell}$ differs from the bulk string operator $O_{\ell}$ introduced in Eq.~\eqref{eq:bulk_string}.
Namely, the original bulk construction $O_{\ell}$ contains a string-like excitation connecting surface anyons while the surface construction $\hat{O}_{\ell}$ does not contain a string portion and hence creates a pair of surface chiral anyons. 
Here, it is useful to explicitly write down the excitation configuration created by the bulk string operator $O_{\ell}$:
\begin{align}
\figbox{0.6}{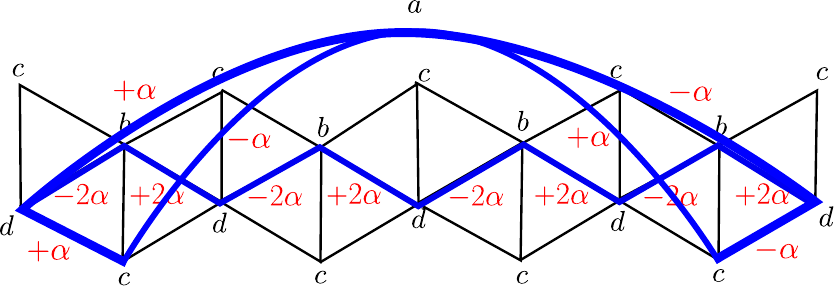}. 
\end{align}
One important observation is that the intermediate portion of the string excitation can be absorbed into the surface:
\begin{align}
\figbox{0.6}{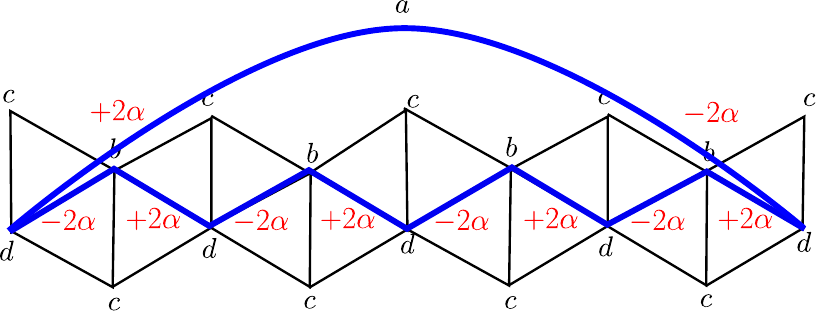} \ 
\end{align}
Combining these operators lead to the correct surface string operator.

Thus, the boundary realizes an anomalous chiral topological order that matches the modular theory $\mathbb{Z}_d^{(\alpha)}$.

%\subsection{Chiral mixed state}

It is interesting to characterize the surface chiral topological order from the perspective of mixed states. Let us begin by defining the boundary stabilizer group. 

\begin{definition}
The boundary stabilizer group $\mathcal{S}^{\partial A}_{\mathrm{chiral-CC}}$ on the $A$ boundary is generated by face stabilizers with color labels $AB,AC,AD$: 
\begin{align}
\mathcal{S}^{\partial_A}_{\mathrm{chiral-CC}} &= \langle X(f_{AB}), X^{-1}Z^{-\alpha}(f_{AC}), Z^{\alpha}(f_{AD}) \rangle,   \qquad f_{AB}, f_{AC}, f_{AD} \in \partial_A \Delta_3 \\
&= \figbox{0.65}{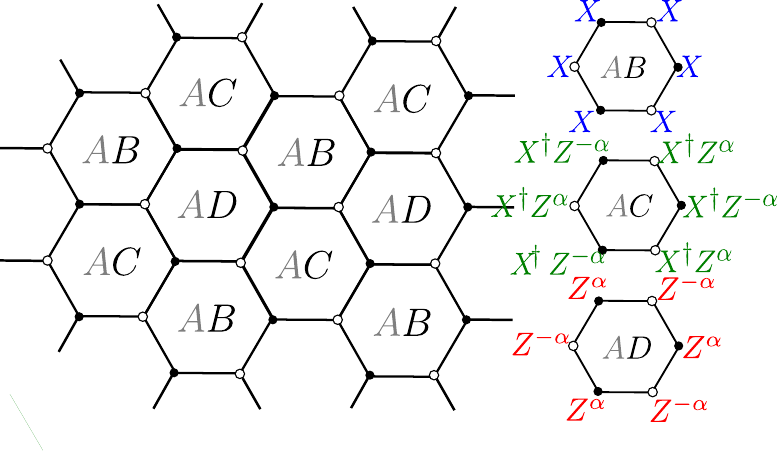} 
\end{align}
where $\partial_A \Delta_3$ denotes the two-dimensional cells on the $A$ boundary of $\Delta_3$. 
\end{definition}

When viewed as a stabilizer code, the boundary code $\mathcal{S}^{\partial A}_{\mathrm{chiral-CC}}$ has code distance $d=2$, due to the following two-body operators that act as logical operators:
\begin{align}
g_{e} \ = \ \figbox{0.65}{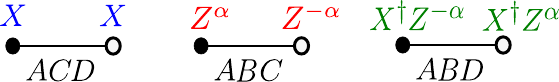} 
\end{align}
Then, by a straightforward extension of the result for the XYZ-color code, we can show that all the codeword states of the boundary code are long-range entangled. 

\begin{claim}
Consider the maximally mixed state $\rho^{\partial_A}_{\mathrm{chiral-CC}}$ that is stabilized by the boundary stabilizer group $\mathcal{S}^{\partial A}_{\mathrm{chiral-CC}}$:
\begin{align}
\rho^{\partial_A}_{\mathrm{chiral-CC}} = \prod_{f \in \partial_A \Delta_3} (I + S_{f})
\end{align}
up to appropriate normalization. Then, $\rho^{\partial_A}_{\mathrm{chiral-CC}}$ cannot be expressed as a convex sum of short-range entangled states. 
\end{claim}

%This claim can be proven by observing that surface string operators possess non-trivial braiding and self statistics. 
%Here, it is crucial that surface string operators are truncated stabilizer operators, which enables us to deform surface string operators, and utilize the argument from~\cite{} to establish the claim. 

\subsection{Top and bottom boundaries}

\begin{figure}
\centering
\raisebox{\height}{\hspace{5pt}}\raisebox{-0.85\height}{\includegraphics[width=0.75\textwidth]{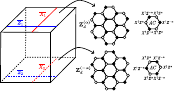}}
\caption{Time-reversal conjugated surface anyon theories on opposite boundaries of the chiral color code.  The bipartite sign structure is reversed across the top and bottom surfaces, leading to $\mathbb{Z}_d^{(\alpha)}$ and $\mathbb{Z}_d^{(-\alpha)}$ surface theories. 
The system encodes $k=2$ logical qudits via anyons on the two boundaries. 
}
\label{fig_chiral_top_bottom}
\end{figure}

When the chiral color code is defined on a thickened torus geometry with two opposite $2$-torus boundaries (top and bottom), the model realizes \emph{time-reversal-conjugate} surface topological orders on opposite boundaries, namely $\mathbb{Z}_d^{(\alpha)}$ and $\mathbb{Z}_d^{(-\alpha)}$. 
This conjugate pair arises from the reversal of the bipartite sign structure on the two boundaries. 
As illustrated in Fig.~\ref{fig_chiral_top_bottom}, black and white vertex assignments are exchanged between the top and bottom surfaces which effectively switches $+\alpha$ to $-\alpha$. 
Since the braiding and self-statistics of the $\mathbb{Z}_d^{(\alpha)}$ theory are complex-conjugated under $\alpha \to -\alpha$ (Eq.~\eqref{eq:braiding}), this sign reversal results in a modular conjugate pair of anyon theories on the opposing surfaces.

This configuration also reflects an emergent \emph{reflection symmetry} of the chiral color code. When the boundary surfaces are given hexagonal lattices, the top and bottom boundaries are related by reflection across a horizontal plane, which exchanges the bipartite sign structure. Thus, in these geometries, time-reversal symmetry is implemented as spatial reflection which exchange the two boundary theories.

A subtle but important point is that the observed chirality depends on the viewing direction. 
In the discussion above, both surfaces are viewed from the same external direction (from above), which reveals the conjugate pair $\mathbb{Z}_d^{(\alpha)}$ and $\mathbb{Z}_d^{(-\alpha)}$. 
If instead one views each boundary theory from the bulk interior, then both appear as $\mathbb{Z}_d^{(\alpha)}$. 
This difference highlights the role of orientation reversal. 
Namely, changing the viewpoint introduces an effective reflection, which flips the sign of the chiral parameter on one surface relative to the other.

\subsubsection*{Implication for MBQC}

Let us now consider the implications for quantum error correction.
In the full 3D gauge color code, defined on the same thickened torus geometry, the code supports no logical qudit.  
However, from the perspective of the stabilizer chiral color code obtained by gauge fixing, the system supports two logical qudits, each encoded on the top and bottom boundaries via the $\mathbb{Z}_d^{(\alpha)}$ and $\mathbb{Z}_d^{(-\alpha)}$ surface anyons. 
As such, a single-shot error-correction procedure prepares a particular codeword state of the chiral color code. 
One can verify that codeword states, prepared by single-shot error correction, are stabilized by 
\begin{align}
\overline{X_1} \ \overline{X_2}, \quad \overline{Z_1}\ \overline{Z_2}
\end{align}
where $\overline{X_1}$, $\overline{Z_1}$ act on the top surface, and $\overline{X_2}$, $\overline{Z_2}$ act on the bottom (Fig.~\ref{fig_chiral_top_bottom}). 
This implies that the two logical qudits are prepared in a maximally entangled EPR pair:
\begin{align}
|\Psi \rangle = \frac{1}{\sqrt{d}} \sum_{j=0}^{d-1} |j\rangle_{\mathrm{top}}\otimes |j\rangle_{\mathrm{bottom}},
\end{align}
demonstrating that the single-shot preparation protocol creates nonlocal entanglement between modular conjugate surface chiral topological codes. 

This maximally entangled state between $\mathbb{Z}_d^{(\alpha)}$ and $\mathbb{Z}_d^{(-\alpha)}$ surfaces suggests that the chiral color code may serve as a resource state for fault-tolerant measurement-based quantum computation (MBQC)~\cite{RBH, Roberts:2016lvf}. 
Namely, a logical qudit encoded in the chiral $\mathbb{Z}_d^{(-\alpha)}$ anyons on the bottom surface can be teleported through the 3D bulk and reconstructed as a conjugate logical qudit on the top surface.
The teleportation is protected against local physical errors, as well as faulty syndromes, due to the single-shot error-correcting property of the 3D gauge color code.
Thus, the chiral color code provides a novel example of single-shot topological teleportation between chiral boundary encodings, mediated by a short-range entangled but non-trivial 3D bulk.

This viewpoint aligns naturally with previous proposals that interpret resource states for topological MBQC in terms of symmetry-protected topological (SPT) phases. 
The standard 3D resource state for MBQC~\cite{RBH} can be understood as a 3D SPT phase protected by two copies of $1$-form $\mathbb{Z}_2$ symmetries~\cite{Yoshida:2015cia}. 
Similarly, the 3D gauge color code—whose codeword subspace is stabilized by volume operators, which possess an intrinsic set of six $1$-form $\mathbb{Z}_d$ symmetries, suggesting a natural SPT interpretation~\cite{Kubica:2018lhn}.
The Walker–Wang models are known to provide classic examples of SPT phases with anomalous surface topological orders, and this connection has already been exploited to show that certain Walker–Wang resource states (e.g., the 3‑fermion state) are universal for MBQC~\cite{Roberts:2020zmk}.
Here, our result hints that MBQC universality is a generic feature of 3D abelian topological order, including chiral topological codes, enabled by single-shot quantum error correction.

\section{Summary and Outlook}

In this paper, we introduced the chiral color codes, a new family of simple three-dimensional stabilizer codes that realize fermionic and chiral topological orders while admitting single-shot error-correction.
Starting with the XYZ color code (a qubit version), we showed that the model realizes the fermionic toric code with anomalous femrionic surface topological order.
We then generalized the model to qudit systems and constructed the chiral color code that capture the topological phases described by $\mathbb{Z}_d^{(\alpha)}$ anyons.
On closed manifolds, we showed that the bulk is short-range entangled for odd $d$ by constructing explicit local quantum channels that prepare the unique ground state.
On open manifolds, we demonstrated that the codes support anomalous chiral surface topological orders, which can be also interpreted as chiral mixed states. 

From a quantum information perspective, our results significantly broaden the landscape of single-shot error-correcting codes and fault-tolerant quantum computation to fermionic and chiral orders. 
From a many-body physics perspective, the simplicity of our stabilizer formulation provides analytical, conceptual, and experimental access to the Walker–Wang type models. 
Also, our results demonstrate that exotic boundary anomalies and bulk short-range entanglement can be captured in minimal stabilizer models, opening the door to systematic exploration of such phases.

Several open directions naturally arise. 

\begin{itemize}
\item The chiral color codes provide another instance where emergent $1$-form symmetries enable single-shot error correction. This suggests a fundamental connection between fault-tolerance, higher-form symmetries, and SPT orders in mixed states.
\item An important challenge is whether stabilizer-like constructions, with the single-shot error-correction capabilities, can capture non-abelian topological phases. 
\item Given that the chiral color codes may serve as conversion platforms between bosonic, fermionic, and chiral degrees of freedom, it is natural to ask whether they imply interesting hybrid quantum computing architectures. 
\end{itemize}

\subsection*{Acknowledgment}

We thank 
Yu-An Chen, 
Tyler Ellison, 
Ryohei Kobayashi, 
Aleksander Kubica, 
Zhi Li, 
Nat Tantivasadakarn, 
Michael Vasmer, 
Chong Wang, 
and
Yijian Zou 
for discussions.
We thank Ray Laflamme for his long-standing support for quantum error-correction researches at Perimeter Institute. 
Research at Perimeter Institute is supported in part by the Government of Canada through the Department of Innovation, Science and Economic Development and by the Province of Ontario through the Ministry of Colleges and Universities. 
This work is supported by the Applied Quantum Computing Challenge Program at the National Research Council of Canada.

\appendix

\section{Condensing bosons}\label{appendix:condensation}

We have demonstrated that the chiral color code corresponds to the 3D topological phase which supports $\mathbb{Z}_d^{(\alpha)}$ surface anyons.
Remarkably, a wide variety of abelian anyon theories can be generated from (possibly multiple copies of) $\mathbb{Z}_d^{(\alpha)}$ theories via boson condensation.
\footnote{Indeed, the Witt class classification seems to suggest that constructions based on $\mathbb{Z}_d^{(\alpha)}$ theories, together with boson condensation, can in principle realize all abelian anyon theories with $\mathbb{Z}_d$ fusion rules.
This observation hints the chiral color code as a particularly useful platform for fault-tolerant realization of arbitrary Abelian anyon theories.
}
Here, we illustrate the procedure of boson condensation for the chiral color code by focusing on the $\mathbb{Z}_4^{(1)}$ cases.
We expect that the procedure is broadly applicable to perform boson condensations in mutliple copies of $\mathbb{Z}_d^{(\alpha)}$ anyons. 

\subsection{Chiral semion from $\mathbb{Z}_4^{(1)}$ }

The $\mathbb{Z}_4^{(1)}$ anyon theory consists of four anyons $\{1,a,a^2,a^3\}$ with the following topological spins and mutual braiding
\begin{align}
\theta(1)=1 , \quad \theta(a)=\omega, \quad \theta(a^2) = 1, \quad \theta(a^3)=\omega, \qquad B(a,a)=\omega^2
\end{align}
where $\omega = i$. 
Here, $a^2$ is a boson, so we can obtain a chiral semion theory $\{1,a\}$ by condensing $a^2$. 
This imposes $a^2=1$ and remove anyons that braid non-trivially with it. 

Boson condensations can be thought of as a projective measurement that immobilizes the bosons~\cite{Ellison_2023}. 
Namely, in the chiral color code, one can condense $a^2$ bosons by projectively measuring the $\mathbb{Z}_4^{(1)}$ model with the following two-body edge operators and post-selecting them to $+1$ eigenstates:
\begin{align}
Z^2(e_{ABC})= \figbox{0.75}{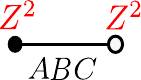}\ ,\quad X^2Z^2 (e_{ABD}) = \figbox{0.75}{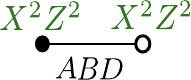}\ , \quad X^2(e_{ACD}) = \figbox{0.75}{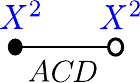} .
\end{align}
Here, it is important to note that these two-body edge operators commute with each other because $a^2$ is a boson with trivial topological spin.

After the projective measurement, the following face stabilizer generators remain:
\begin{align}
X(f_{AB}),\quad X^{-1}Z^{-1}(f_{AC}), \quad Z(f_{AD}), \quad X^2(f_{CD}),\quad X^{2}Z^{2}(f_{BD}), \quad Z^2(f_{BC}). 
\end{align}
Namely, $X(f_{CD})$, $X^{-1}Z^{-1}(f_{BD})$, and $Z(f_{BC})$ are no longer stabilizers as they do not commute with measured edge operators.

The surface topological order after condensation is characterized by the boundary stabilizer code:
\begin{align}
\mathcal{S}^{\partial_A}_{\mathrm{semion-CC}} &= \langle X(f_{AB}), X^{-1}Z^{-1}(f_{AC}), Z(f_{AD}) , X^2(e_{ACD}), X^2Z^2 (e_{ABD}),  Z^2(e_{ABC})  \rangle  \\
&= \figbox{0.65}{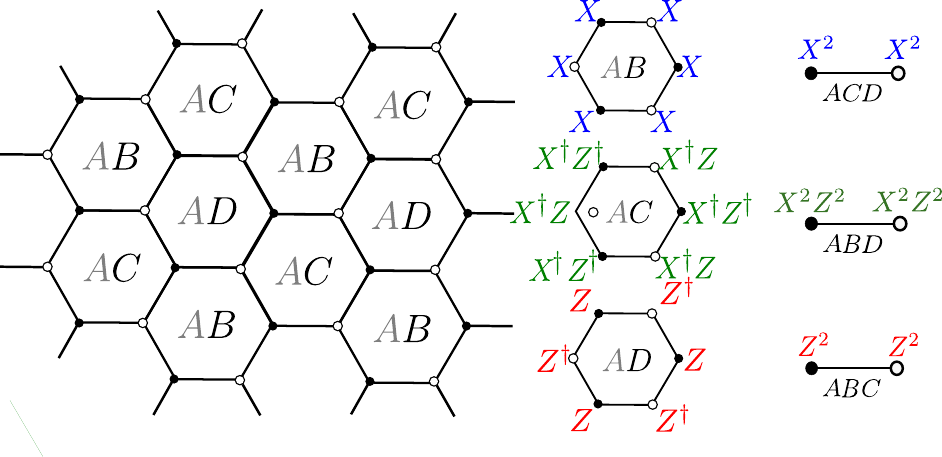} \ .
\end{align}
We notice that the surface string operator $\hat{O}_{\ell}$ creates chiral semions while its square $\hat{O}_{\ell}^2$ decomposes into a product of edge operators, confirming that the surface anyon theory is indeed chiral semion with bosons $a^2$ condensed. 

Because the bosonic $\mathbb{Z}_2$ bulk charge has been removed by condensation, the bulk becomes short-range entangled.
In the original $\mathbb{Z}_4^{(1)}$ chiral color code, applying $O_{\ell}^2$ created a pair of bulk charges.
In the condensed code, however, these charges no longer exist since the new face stabilizers, $X^2(e_{ACD}), X^2Z^2 (e_{ABD}),  Z^2(e_{ABC})$, commute with $O_{\ell}^2$.
As a result, the ground state of the chiral semion color code can be prepared by a local quantum channel using the same local error-correction strategy as before.

\subsection{Three-fermion theory from $\mathbb{Z}_4^{(1)}\times \mathbb{Z}_4^{(1)}\times \mathbb{Z}_4^{(1)}\times \mathbb{Z}_4^{(1)} $ }

Next, consider four copies of chiral semion theories 
\begin{align}
\{ 1, a_1 \} \times \{ 1, a_2 \} \times \{ 1, a_3 \} \times \{ 1, a_4 \}
\end{align}
where we imposed $a_j^2=1$ in each copy. 
The composite $a_1 a_2 a_3 a_4$ is a boson, and condensing it produces the three-fermion theory with anyons:
\begin{align}
\{ 1, a_1 a_2, a_1 a_3, a_1 a_4 \}
\end{align}
whose topological spins and mutual braiding are given by
\begin{align}
\theta (a_1 a_2) = \theta(a_1 a_3) = \theta( a_1 a_4 ) = -1,\ \ 
B(a_1 a_2, a_1 a_3) = B(a_1 a_2, a_1 a_4) = B(a_1 a_3, a_1 a_4) = -1.
\end{align}

In order to obtain the three-fermion theory, one needs to condense $a_j^2$ as well as $a_1 a_2 a_3 a_4$ by some projective measurements. 
Unlike the $\mathbb{Z}_4^{(1)}$ case, measuring two-body edge operators do not work as some of the two-body edge operators do not commute with each other.
Namely, consider the following edge operators corresponding to $a_1 a_2 a_3 a_4$:
\begin{align}
\bigotimes_{j=1}^4 Z_{j}(e_{ABC}), \qquad 
\bigotimes_{j=1}^4 X^{-1}Z^{-1}_{j}(e_{ABD}), \qquad 
\bigotimes_{j=1}^4 X_{j}(e_{ACD}).
\end{align}
We then notice that these edge operators do not commute with edge operators for $a_j^2$. 

This issue can be resolved by multiplying two edge operators to form three-body (short) string operators. Namely, to condense $a_j^2$, we measure short strings that start from white dots and end on white dots. An example is shown below:
\begin{align}
X^2(e_{ACD})X^2Z^2(e_{ABD}) =  \figbox{0.65}{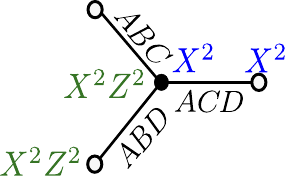}
\end{align}
On contrary, to condense $a_1 a_2 a_3 a_4$, we measure short strings that start from black dots and end on black dots:
\begin{align}
\bigotimes_{j=1}^4 X_j(e_{ACD})X_j^{-1}Z_j^{-1}(e_{ABD}) = \bigotimes_{j=1}^4  \figbox{0.65}{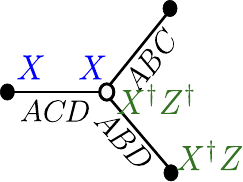}
\end{align}
One can then verify that all these short string operators commute with each other, and thus can be simultaneously measured. 
This way, one can obtain the three-fermion theory. 
The fact that these short strings commute with each other is closely related to the trivial mutual braiding between $a_1 a_2 a_3 a_4$ and $a_j$.
Namely, one can verify that the group commutator of short strings corresponds to the mutual braiding statistics.

As these two examples suggest, boson condensations in the chiral color code can be performed by measuring short string operators corresponding to the bosons which are to be condensed. 
In the simplest case of $\mathbb{Z}_4^{(1)}$, measuring two-body edge operators suffice. 
In the case of multiple copies, measuring products of two-body edge operators (so, three-body operators) lead to appropriate boson condensations. 
We hope to formalize the condensation procedure and the recipe to create arbitrary abelian anyon theories within the framework of the chiral color code in the future work.

%The above procedures illustrate that boson condensation can be implemented systematically within the chiral color code by local projective measurements of edge operators.
%This approach is not limited to the chiral semion and three-fermion theories. 
%Namely, by stacking multiple copies of the $\mathbb{Z}_d^{(\alpha)}$ chiral color code and condensing appropriate bosons via projective measurement of edge operators, any abelian anyon theories may be realized.
%In this sense, the chiral color code provides a unified and fault-tolerant framework for constructing a wide variety of abelian topological orders, including those with chiral or fermionic boundary excitations.

%===============================================================================
\mciteSetMidEndSepPunct{}{\ifmciteBstWouldAddEndPunct.\else\fi}{\relax}
\bibliographystyle{JHEP}
\bibliography{references.bib}

\end{document}